\documentclass[fleqn,usenatbib]{mnras}

\usepackage{newtxtext,newtxmath}

\usepackage[T1]{fontenc}
\usepackage{ae,aecompl}

\usepackage{epsfig,graphicx}	
\usepackage{amsmath}	
\usepackage{natbib}
\usepackage{bm}
\usepackage{comment}

\title[MAXI J1820+070 Jet Variability]
{Measuring fundamental jet properties with multi-wavelength fast timing of the black hole X-ray binary  MAXI J1820+070}

\author[A.J. Tetarenko et al.]{A.J. Tetarenko$^{1}$\thanks{E-mail: a.tetarenko@eaobservatory.org},
P. Casella$^2$,
J.C.A. Miller-Jones$^3$,
G.R. Sivakoff$^4$,
J.A. Paice$^5$,
\newauthor
F.M. Vincentelli$^5$,
T.J. Maccarone$^6$,
P. Gandhi$^5$,
V.S. Dhillon$^{7,8}$,
T.R. Marsh$^9$,
\newauthor
T.D. Russell$^{10,11}$,
and P. Uttley$^{11}$
\\
$^1$East Asian Observatory, 660 N. A'oh\={o}k\={u} Place, University
Park, Hilo, Hawaii 96720, USA\\
$^2$INAF-Osservatorio Astronomico di Roma, Via Frascati 33, I-00078 Monteporzio Catone, Italy\\
$^3$International Centre for Radio Astronomy Research - Curtin University, GPO Box U1987, Perth, WA 6845, Australia\\
$^4$Department of Physics, University of Alberta, CCIS 4-181, Edmonton, AB T6G 2E1, Canada\\
$^5$School of Physics and Astronomy, University of Southampton, Southampton, SO17 1BJ, UK\\
$^6$Department of Physics and Astronomy, Texas Tech University, Lubbock, Texas 79409-1051, USA\\
$^7$Department of Physics and Astronomy, University of Sheffield, Sheffield, S3 7RH, UK\\
$^8$Instituto de Astrofisica de Canarias, E-38205 La Laguna, Tenerife, Spain\\
$^9$Astronomy and Astrophysics Group, Department of Physics, University of Warwick, Gibbet Hill Road, Coventry, CV4 7AL, UK\\
$^{10}$INAF-Istituto di Astrofisica Spaziale e Fisica Cosmica, Via U. La Malfa 153, I-90146 Palermo, Italy\\
$^{11}$Anton Pannekoek Institute for Astronomy, University of Amsterdam, Science Park 904, 1098 XH Amsterdam, The Netherlands\\
}

\date{Accepted XXX. Received YYY; in original form ZZZ}

\pubyear{2021}

\begin{document}
\label{firstpage}
\pagerange{\pageref{firstpage}--\pageref{lastpage}}
\maketitle

\begin{abstract}
We present multi-wavelength fast timing observations of the black hole X-ray binary MAXI J1820+070 (ASASSN-18ey), taken with the Karl G. Jansky Very Large Array (VLA), Atacama Large Millimeter/Sub-Millimeter Array (ALMA), Very Large Telescope (VLT), New Technology Telescope (NTT), Neutron Star Interior Composition Explorer (NICER), and XMM-Newton. Our data set simultaneously samples ten different electromagnetic bands (radio -- X-ray) over a 7-hour period during the hard state of the 2018--2019 outburst. The emission we observe is highly variable, displaying multiple rapid flaring episodes. To characterize the variability properties in our data, we implemented a combination of cross-correlation and Fourier analyses. We find that the emission is highly correlated between different bands, measuring time-lags ranging from hundreds of milliseconds between the X-ray/optical bands to minutes between the radio/sub-mm bands. Our Fourier analysis also revealed, for the first time in a black hole X-ray binary, an evolving power spectral shape with electromagnetic frequency. Through modelling these variability properties, we find that MAXI J1820+070 launches a highly relativistic ($\Gamma=6.81^{+1.06}_{-1.15}$) and confined ($\phi=0.45^{+0.13}_{-0.11}$ deg) jet, which is carrying a significant amount of power away from the system (equivalent to $\sim0.6 \, L_{1-100{\rm keV}}$). We additionally place constraints on the jet composition and magnetic field strength in the innermost jet base region.
Overall, this work demonstrates that time-domain analysis is a powerful diagnostic tool for probing jet physics, where we can accurately measure jet properties with time-domain measurements alone.

\end{abstract}



\section{Introduction}
\label{sec:intro}
One of the key open questions in high energy astrophysics is the nature of the connection between accretion and relativistic jets launched from compact objects. Determining how jets arise as a result of accretion, and quantifying how much energy they inject into the local environment, are important problems, as jet feedback can affect many other astrophysical processes (such as star formation and galaxy evolution; \citealt{silk98,mir11}). However, we still do not understand the complex relationship between the properties of the accretion flow (geometry, mass accretion rate), and the properties of relativistic jets (kinetic power, bulk speed). 

A crucial step towards understanding the inner workings of these jets is characterizing jet properties and how they evolve with the accretion flow.  Black hole X-ray binaries (BHXBs) are ideal test-beds to study jets, as they are close in proximity, display a wide range of accretion and jet launching environments, and evolve on human timescales. BHXBs contain a stellar-mass black hole accreting matter from a companion star, where a portion of the accreted material can be ejected in the form of a relativistic jet.
{The majority of these systems are transient, evolving from periods of minimal activity into bright outbursts (over timescales of days to months), during which the system evolves through several different accretion states \citep{fbg04,remmc06,fenhombel09,bel10,tetarenkob2015,cs16}}. 
{Compact jet emission observed in the hard accretion state of a BHXB outburst} is produced as a result of synchrotron radiation \citep{hj88aa,corfen02}, and displays a characteristic spectrum consisting of an inverted optically thick portion ($\alpha>0$, where flux density scales as $\nu^\alpha$), which breaks to an optically thin ($\alpha\sim-0.7$) portion (this synchrotron spectral break is thought to mark the most compact jet base region where particle acceleration begins; \citealt{mar01,mar05,rus12,rus13}). Therefore, jet synchrotron emission tends to dominate in the lower electromagnetic frequency bands (radio, sub-mm, infrared, and possibly optical; \citealt{fen01,rus06,tetarenkoa2015}), while emission from the accretion flow dominates in the higher electromagnetic frequency bands (optical, UV, X-ray; \citealt{d07}).

The jet launched during a BHXB outburst is known to produce highly-variable emission, and thus time-domain analyses can offer a new avenue to probe detailed jet properties \citep{cas10,gan17,teta19}. For instance, Fourier domain measurements can probe physical scales not accessible with current imaging capabilities (sub-AU scales at kpc distances). This allows us to map out the jet size scale with respect to the distance downstream from the black hole, and in turn make geometric measurements of the jet cross-section and opening angle (in the case of a conical jet).
Further, due to optical depth effects, emission originating from a BHXB jet displays a distinct observational signature where the signal at lower electromagnetic frequencies will appear as a delayed version of the signal at higher electromagnetic frequencies.
Through combining size scale constraints at different electromagnetic frequencies, with time-lag measurements between emission features (e.g., flares) at these electromagnetic frequencies, we are able to place constraints on jet power and speed (also possibly jet acceleration over different scales; \citealt{blandford79,heinz06,utt15}).
Lastly, with simultaneous radio, sub-mm, infrared/optical (OIR), and X-ray observations we are also able to cross-correlate multi-band jet/accretion flow variability signals. This type of analysis allows us to directly link changes in the accretion flow (probed by X-ray variability) with changes in the jet on different scales (OIR probes the jet base, radio probes further along the flow).

Multi-wavelength spectral timing studies on BHXBs were first performed in the optical bands (e.g., \citealt{motch82,kan01,ganh08}), but it was only recently that the first unambiguous detection of (stationary\footnote{Alternatively, see for example \citealt{fend98} studying IR quasi-periodic oscillations detected in the BHXB GRS 1915+105.}) IR variability \citep{cas10,vinc18} originating in a compact jet came from GX 339--4 (as optical frequencies can at times contain contributions from other emission sources, such as the accretion disc). These works discovered that IR emission originating from the base of the jet varied significantly on sub-second timescales, IR variability was highly correlated with X-ray variability, and the IR jet variability properties (i.e., time-lags, variability amplitude) appeared to be vastly different at different stages of the outburst \citep{kalam16}. With similar sub-second correlated variability signatures now also seen in several outbursting sources in the optical \citep{gan10, gan17, pai19}, there is little doubt that a common mechanism must be driving this variability. 
Further, \citet{mal18} have recently shown that IR variability properties measured in the Fourier domain (e.g., power spectra, coherence, lags) can be reproduced by a jet model where the variability is driven by internal shocks in the jet flow created through the collision of discrete shells of plasma injected at the base of the jet with variable speeds (the behaviour of these shells is directly linked to the amplitude of X-ray variability at different timescales; see also \citealt{jam10,mal14,drap15,drap17,pea19,bassi20,marino20}). The same model can also explain broadband IR spectral properties, including the presence of the synchrotron spectral break, which under the assumption of a single-zone model is an estimator of the magnetic field strength in the jet plasma and the size of the emitting region in this first acceleration zone at the base of the jet \citep{chat,gan11}.
This new work opens up the possibility of studying internal processes in the jet through measuring variability properties in the Fourier domain.
Overall, all of these results suggest that variability in the accretion flow is subsequently driving variability in the jet, and confirmed the diagnostic potential of time domain studies for studying jet and accretion physics in BHXBs.

Recently, \citet{teta19} expanded these BHXB spectral timing studies into the radio frequency bands (probing further out along the jet flow when compared to OIR frequencies), by repeating this spectral timing experiment using simultaneous radio and X-ray observations of the BHXB Cygnus X--1. Due to the vastly different instruments and observing techniques used at radio frequencies (when compared to OIR frequencies), to enable a radio timing study the authors implemented a new technique whereby a full interferometric array was split into separate sub-arrays, allowing for the periods of continuous, multi-band data needed to use Fourier and cross-correlation analyses to study the variability.
This work was able to connect rapid variability properties in radio light curves to real jet physics in a BHXB for the first time. In particular, the jet speed and opening angle were measured through detecting and modelling time-lags between the radio/X-ray signals at different bands, and a Fourier analysis of the radio signals further allowed the authors to track how matter propagates through the radio emission regions, revealing new information about the internal processes occurring in the jet.
These new results show that not only is time-domain analysis possible in the lower electromagnetic frequency bands, but it is a powerful tool that can provide an unprecedented view of a BHXB jet.

In this work, we build on the success of this first radio spectral timing study by using the sub-array technique to observe the BHXB MAXI J1820+070 during its 2018 -- 2019 outburst. Here we further improve upon the Cygnus X--1 experiment by adding sub-mm, infrared, and optical observations to sample more electromagnetic frequency bands, in turn allowing us to connect variability properties across different scales in the jet (from the jet base probed by the infrared/optical, to regions further out along the jet flow probed by the radio/sub-mm).

\subsection{MAXI J1820+070}
\label{sec:m1820}
MAXI J1820+070 (also known as ASASSN-18ey) is a dynamically confirmed, low-mass BHXB, containing a K-type dwarf donor star and a $8.48^{+0.79}_{-0.72}M_\odot$ black hole \citep{torr19,torr20}. This source was first discovered as a bright optical transient by the All-Sky Automated Survey for Supernovae (ASASSN; \citealt{kochan17}) when it entered into outburst in 2018 March \citep{denis18,tuck18}. Archival photographic plates\footnote{Digital Access to a Sky Century at Harvard (DASCH; \url{http://dasch.rc.fas.harvard.edu}).} have since presented evidence of two past outbursts of this system, in 1898 and 1934 \citep{koj19}. 
An X-ray counterpart to the optical source was subsequently detected by the Monitor for All-sky X-ray Image (MAXI; \citealt{kawa18,shid18}), while a radio counterpart was detected with the Arcminute Microkelvin Imager Large Array (AMI-LA; \citealt{bri18}) and the RATAN-600 radio telescope \citep{trus18}. The radio source was shown to have a flat spectrum extending up to the sub-mm bands \citep{teta18m1820a}, consistent with the presence of a partially self-absorbed compact jet. 
MAXI J1820+070 remained in a hard accretion state, with a compact jet present, from 2018 March until early 2018 July \citep{homa18}, when it began transitioning to the soft accretion state. At this point, the compact jet was quenched \citep{teta18m1820b}, and superluminal discrete jet ejections were launched \citep{bright20,esp20}.
Recent radio parallax measurements indicate that MAXI J1820+070 is located at a distance of $2.96\pm0.33$ kpc, and the jets have an inclination angle of $63\pm3$ degrees\footnote{This inclination angle estimate was derived using the combination of jet proper motions and the known distance.} to our line of sight \citep{atri20}. The recently released Gaia EDR3 \citep{gaiaedr3} distance of $2.66^{+0.85}_{-0.52}$ kpc (calculated using the recommended zero-point correction and a prior that models the XB density distribution in the Milky Way; \citealt{grimm2002,atri19,lind20}) is consistent with this radio parallax estimate, improving upon the larger uncertainties reported in DR2 \citep{gand18}.

MAXI J1820+070 has been found to be highly variable down to sub-second timescales in the X-ray and optical bands \citep{kaj19,pai19,mud20,st20,buis21}. In this work, we focus on connecting variability from across the electromagnetic spectrum, combining fast timing observations at radio, sub-mm, infrared, optical, and X-ray frequencies. In \S\ref{sec:data}, we outline the data collection and reduction processes. In \S\ref{sec:measure}, we present high time resolution light curves of MAXI J1820+070, and characterize the variability we observe in these light curves with Fourier domain and cross-correlation analyses. In \S\ref{sec:model}, we model the jet timing properties derived from our Fourier and cross-correlation analyses, placing constraints on jet power, speed, geometry, and size scales. In \S\ref{sec:discuss}, we discuss the time-domain properties of the jet in MAXI J1820+070, and highlight important considerations in designing future spectral timing experiments of BHXB jets. A summary of the results is presented in \S\ref{sec:sum}.

\section{Observations and data analysis}
\label{sec:data}

\subsection{ALMA sub-mm observations}
\label{sec:almasub}

MAXI J1820+070 was observed with the Atacama Large Millimeter/Sub-Millimeter Array (ALMA; Project Code: 2017.1.01103.T) on 2018 April 12, for a total on-source observation time of 2.0 hours. Data were taken in Band 7, at a central frequency of 343.5\,GHz. The ALMA correlator was set up to yield $4\times2$ GHz wide base-bands, with a 2.0-second correlator dump time. During our observations, the array was in its Cycle 5 C3 configuration, with 46 antennas. We reduced and imaged the data within the Common Astronomy Software Application package (\textsc{casa} v5.4; \citealt{mc07}), using standard procedures outlined in the \textsc{casa}Guides for ALMA data reduction\footnote{\url{https://casaguides.nrao.edu/index.php/ALMAguides}}. We used J1751+0939 as a bandpass/flux calibrator, and J1830+0619 as a phase calibrator. To obtain high time resolution flux density measurements, we used our custom \textsc{casa} variability measurement scripts\footnote{\url{https://github.com/Astroua/AstroCompute_Scripts}}. These scripts perform multi-frequency synthesis imaging for each time-bin with the \texttt{tclean} task, using a natural weighting scheme to maximize sensitivity. Flux densities of the source in each time bin are then measured by fitting a point source in the image plane (with the \texttt{imfit} task). We performed phase-only self-calibration (down to a solution interval of 60 sec) prior to running the imaging scripts. To check that any variability observed in these MAXI J1820+070 sub-mm frequency light curves is dominated by intrinsic variations in the source, and not due to atmospheric or instrumental effects, we also ran additional tests on a check source (see Appendix~\ref{sec:ap_cal} for details). An time-averaged flux density measurement over the whole observation period is shown in Table~\ref{table:avgfluxes}.

   \begin{figure*}
\begin{center}
  \includegraphics[width=1\textwidth]{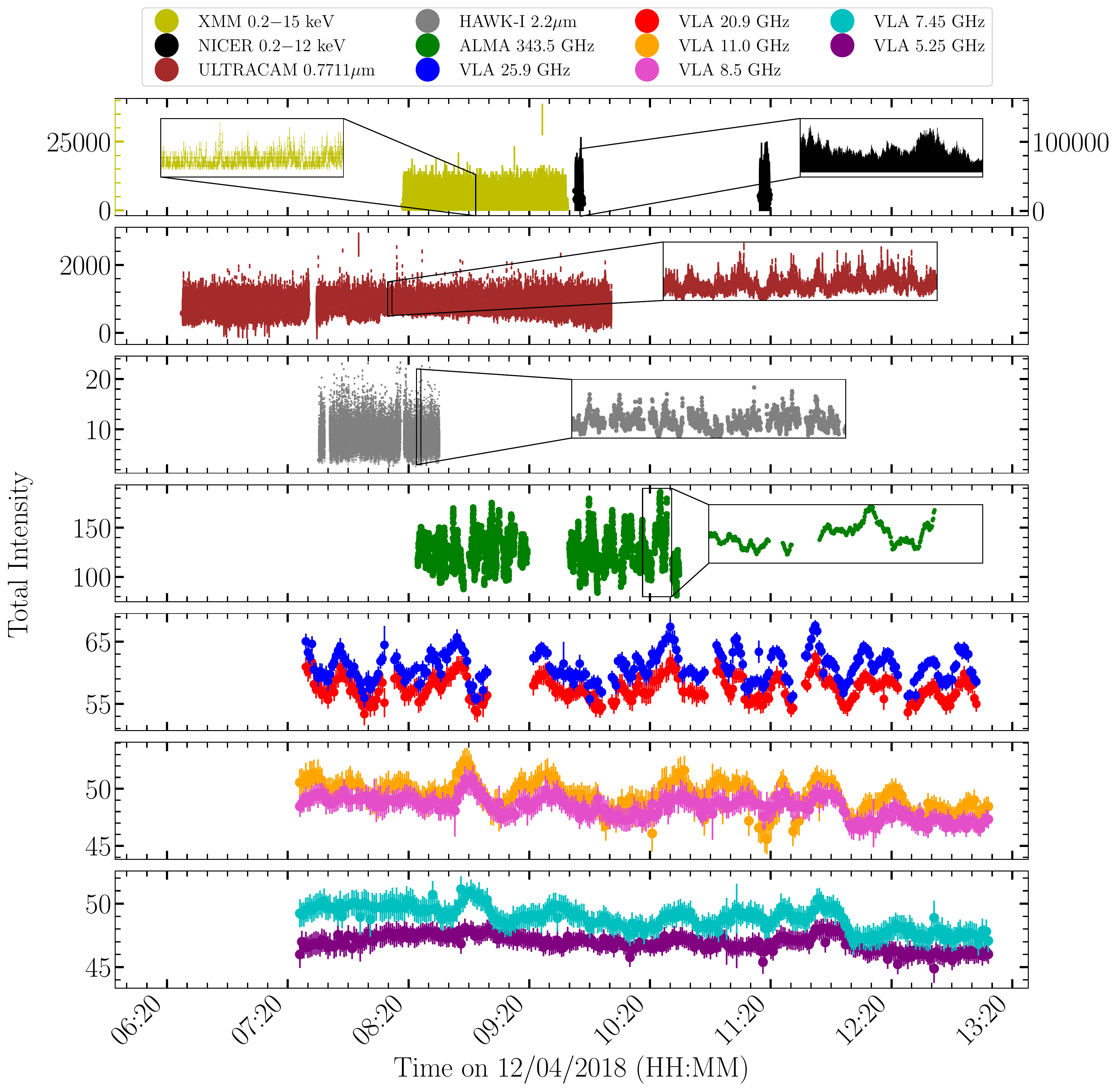}\\
 \caption{\label{fig:lc}Simultaneous multi-band light curves of the BHXB MAXI J1820+070 taken on 2018 April 12 in the X-ray (XMM-Newton 0.2--15 keV and NICER 0.2--12 keV), optical (0.7711$\mu$m), infrared (2.2$\mu$m), sub-mm (343.5 GHz), and radio (5.25--25.9 GHz) bands. The main panels from \textit{top} to \textit{bottom} show light curves for progressively decreasing electromagnetic frequency bands, and the inset panels show zoomed in versions of the light curves. The total intensity units for VLA/ALMA are mJy/bm, HAWK-I is amplitude with respect to the reference star, ULTRACAM is arbitrary instrumental units, and NICER/XMM-Newton are counts/sec. {The time resolution for the VLA, ALMA, HAWK-I, ULTRACAM, NICER, and XMM-Newton light curves are: 5 sec, 2 sec, 0.0625 sec, 0.01 sec, 0.01 sec, and 0.004 sec, respectively.} We observe clear variability in the emission from MAXI J1820+070, taking the form of rapid flaring across all of the electromagnetic bands sampled. 
 }
\end{center}
 \end{figure*}
 
\begin{figure}
\begin{center}
  \includegraphics[width=0.45\textwidth]{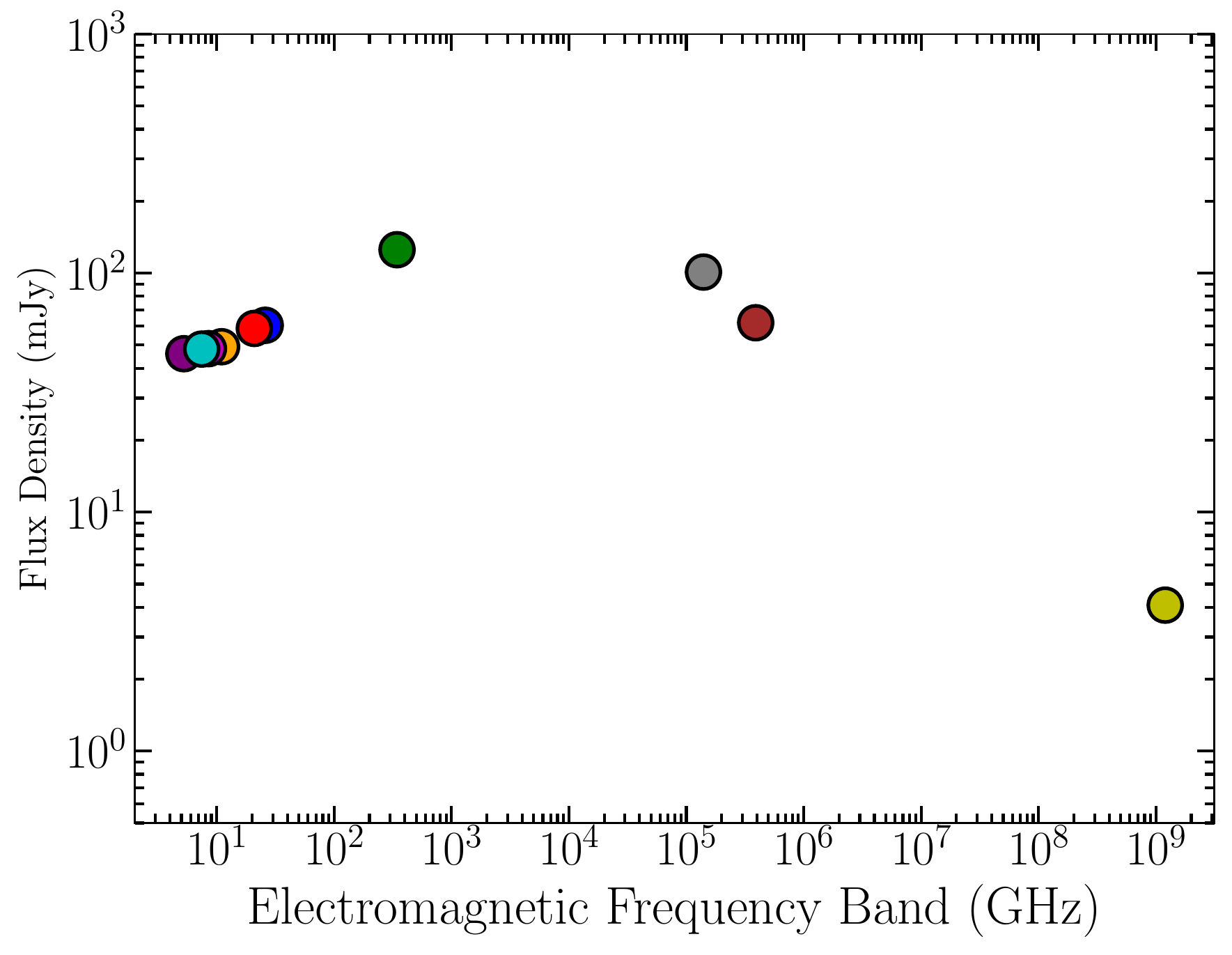}\\
 \caption{\label{fig:sed} Time-averaged broad-band spectrum of our radio--X-ray data of MAXI J1820+070 (see Table~\ref{table:avgfluxes} and \S\ref{sec:infrared}, \ref{sec:opt}, \ref{sec:xray}). The colours of the data points correspond to the same colours of the electromagnetic frequency bands in Figure~\ref{fig:lc}. The radio through sub-mm data appear to lie on the slightly inverted optically thick portion of the jet spectrum (spectral index $\alpha_{\rm thick}\sim 0.25$), while the infrared and optical data appear to lie on the steep optically thin portion of the jet spectrum (spectral index $\alpha_{\rm thin}\sim -0.5$).}
\end{center}
 \end{figure}
 
\subsection{VLA radio observations}
\label{sec:vlarad}
MAXI J1820+070 was observed with the Karl G. Jansky Very Large Array (VLA; Project Code: 18A-470) on 2018 April 12, for a total on-source observation time of 6.0 hours. The array was in the A configuration at the time of our observations, where we split the full array into 3 sub-arrays of 10, 9, and 8 antennas. Observations in each sub-array were made with the 8-bit samplers, where each sub-array observed exclusively in one frequency band; C (4 -- 8 GHz), X (8 -- 12 GHz), or K (18 -- 26 GHz) band. Each band was comprised of 2 base-bands, with 8 spectral windows of 64 2-MHz channels each, giving a total bandwidth of 1.024 GHz per base-band. The sub-array setup allows us to push to shorter correlator dump times than would be possible if we were using the full array. In these observations, we set a 0.15-second correlator dump time, providing the highest time resolution possible, while staying within the standard 25 ${\rm Mb\, s}^{-1}$ data rate limit. With our sub-array setup, we record data fast enough to probe timescales down to hundreds of ms. We implemented a custom non-periodic target/calibrator cycle for each sub-array, alternating observing MAXI J1820+070 and calibrators, at one band per sub-array, such that we obtained simultaneous data in all three bands.  We reduced and imaged the data within \textsc{casa}, using standard procedures outlined in the \textsc{casa}Guides\footnote{\url{https://casaguides.nrao.edu/index.php/Karl\_G.\_Jansky_VLA_Tutorials}.} for VLA data reduction (i.e., a priori flagging, setting the flux density scale, initial phase calibration, solving for antenna-based delays, bandpass calibration, gain calibration, scaling the amplitude gains, and final target flagging). We used 3C286 (J1331+305) as a flux/bandpass calibrator, and J1824+1044 as a phase calibrator. To obtain high time resolution flux density measurements, we follow the same procedure as for the ALMA data, using our custom \textsc{casa} variability measurement scripts. We performed phase-only self-calibration (down to a solution interval of 10 sec) prior to running the imaging scripts. To check that any variability observed in these MAXI J1820+070 radio frequency light curves is dominated by intrinsic variations in the source, and not due to atmospheric or instrumental effects, we also ran additional tests on a check source (see Appendix~\ref{sec:ap_cal} for details).
An time-averaged flux density measurement over the whole observation period is shown for each band in Table~\ref{table:avgfluxes}.

\subsection{VLT HAWK-I infrared observations}
\label{sec:infrared}

HAWK-I is a wide-field photometer operating in the near-IR band (0.97 to 2.31 $\mu$m) made by four HAWAII 2RG 2048x2048 pixel detectors \citep{Pirard2004}. High time resolution near-IR  ($K_s$ band; $2.2\mu$m; $1.4\times10^5$ GHz) data of MAXI J1820+070 was collected on 2018 April 12 between 07:32:58 -- 08:31:39 UTC with the HAWK-I instrument mounted on the Very Large Telescope (VLT; Project Code: 0100.D-0308(A)) UT-4/Yepun. The instrument was set up  in \textit{Fast-Phot} mode: i.e. the detector was limited only to a stripe made of 16 contiguous windows of 64\,$\times$\,64 pixels in each quadrant, permitting a time resolution of {0.0625 sec}. In order to read out the data, the final light curve has periodic gaps of $\approx$ 3 sec every 15--16 sec.
We pointed the instrument in order to place  the target and a bright reference star (K$_s$= 11.9)
in the lower-left quadrant (Q1). We extracted photometric data using the ULTRACAM data reduction pipeline v9.14 tools\footnote{\url{http://deneb.astro.warwick.ac.uk/phsaap/ultracam/}} \citep{dhilon2007}, deriving the parameters from the bright reference star and its position (the position of the target was linked in each exposure). To avoid seeing effects, we used the ratio between the source and the reference star count rate. Finally, we put the time of each frame in the Barycentric Dynamical Time (BJD\_TDB) system. We also derive a rough estimate of the time-averaged flux density from these data, to help put our longer wavelength measurements into context, by comparing the target to the reference star. With $E(B-V)=0.18$ \citep{tuck18}, this leads to a de-reddened flux density of $\sim101$ mJy.

\subsection{NTT ULTRACAM optical observations}
\label{sec:opt}

ULTRACAM is a fast-timing optical instrument which uses three channels for simultaneous multi-wavelength monitoring, and can observe at a high frame rate \citep{dhilon2007}.
High time resolution optical ($i_s$ band\footnote{Please note that $u_s$ and $g_s$ band data were also obtained from these observations. These data will be reported in a separate paper; Paice et al., in prep.}; $0.7711\mu$m; $3.9\times10^5$ GHz) data of MAXI J1820+070 was collected on 2018 April 12 between 06:28 -- 10:01 UTC with the ULTRACAM instrument mounted on the 3.5\,m New Technology Telescope in La Silla, Chile (NTT; Project Code: 0101.D-0767). The instrument was used in two-window mode (one each for the target and the comparison star), with both window sizes of 50\,$\times$\,50 pixels with a 2\,$\times$\,2 binning for sensitivity, and speed at 96.5\,Hz in $i_s$ band
(using ULTRACAM's `co-adding' feature, $i_s$ band was observed with an exposure time of 8.96\,ms, and a total cycle time of 10.4\,ms, giving 1.44\,ms of dead time and a sampling rate of $\sim$96.5\,Hz). 

The data were reduced using the ULTRACAM pipeline v9.14 \citep{dhilon2007}. The bias was subtracted from each frame, and flat field corrections were also applied. Aperture sizes scaled to the instantaneous seeing were used, with radii between 0.7$\arcsec$ and 3.5$\arcsec$, with an annulus of between 12$\arcsec$ and 6.3$\arcsec$ to calculate the background. These apertures had variable centre positions that tracked the centroids of the sources on each frame, with a two-pass iteration (where an initial pass is made to track the sources on the CCD before a second photometry pass) used for accuracy. Our times were then adjusted to BJD\_TDB system using methods given in \cite{eastman_achieving_2010}. 

Our comparison star 
is listed as PSO J182026.430+071011.742 in the \textsc{PanSTARRS} survey catalog \citep{Magnier_PanSTARRS_2020}. 
We extracted the target count rates and count rates for the comparison star using aperture photometry with a variable aperture size dictated by the seeing conditions. The aperture also tracked the centroid of the source of interest by using a bright star in the field as a reference source. 
We also derive a rough estimate of the time-averaged flux density from these data, to help put our longer wavelength measurements into context, by comparing the target to the reference star. With $E(B-V)=0.18$ \citep{tuck18}, this leads to a de-reddened flux density of $\sim62$ mJy.

\subsection{NICER and XMM-Newton X-ray observations}
\label{sec:xray}

 \renewcommand\tabcolsep{6pt}
 \begin{table}
\caption{Time-averaged radio and sub-mm flux densities of MAXI J1820+070}\quad
\centering
\begin{tabular}{ lcc}
 \hline\hline
 {\bf Electromagnetic}&{\bf Average Flux}&{\bf Systematic}\\
  {\bf Frequency Band}&{\bf Density (mJy)$^\dagger$}&{\bf Error (\%)$^\star$}\\[0.15cm]
  \hline
  5.25 GHz&$46.0\pm0.1$&5\\[0.1cm]
  7.45 GHz&$48.1\pm0.2$&5\\[0.1cm]
  8.5 GHz&$48.3\pm0.2$&5\\[0.1cm]
  11.0 GHz&$49.2\pm0.2$&5\\[0.1cm]
  20.9 GHz&$58.7\pm1.1$&10\\[0.1cm]
  25.9 GHz&$60.5\pm1.1$&10\\[0.1cm]
  343.5 GHz&$125.3\pm0.05$&5\\[0.1cm]
 \hline
\end{tabular}\\
\begin{flushleft}
$^\dagger$ Errors on the VLA and ALMA average flux density measurements are calculated from the local rms in images made from data taken over the entire observation period.\\
$^\star$ Additional systematic errors added (in quadrature) to the average flux density error measurements, to account for absolute flux calibration uncertainties at the VLA and ALMA.
\end{flushleft}
\label{table:avgfluxes}
\end{table}
\renewcommand\tabcolsep{6pt} 

NICER (Neutron star Interior Composition ExploreR) is an X-ray instrument aboard the International Space Station (ISS). It comprises 52 functioning X-ray concentrator optics and silicon drift detector pairs, arranged in seven groups of eight. Individual photons between 0.2--12 keV, and their energies, can be detected to a time resolution of 40\,ns \citep{GendreauArzoumanian_NICER_2016}.

For this work, we reduced NICER data from ObsID 1200120127 using {\sc nicerdas}, a collection of NICER-specific tools which is apart of the {\sc heasarc} software package\footnote{\url{https://heasarc.gsfc.nasa.gov}}. Full Level2 calibration and screening was conducted with the {\texttt {nicerl2}} task, which calibrated, checked the time intervals, merged, and cleaned the data. Barycentric correction was carried out using {\sc barycorr}.

{XMM-Newton observed MAXI J1820+070 with the EPIC-pn camera for approximately 2 hours, between 07:39:28 -- 9:39:28 UTC on 2018 April 12. These observations were set up in \textit{Burst Mode}. Events were extracted with a  PATTERN <= 4, FLAG==), in the 0.2--15 keV energy range. We used a box of angular size $\approx$ 86 arcsec (RAWX between 28 and 48). The event file was barycentred using the command \texttt{barycen}, and the XMM-Newton X-ray light curve was extracted with 1 ms time resolution. Due to the very high count rate, the source showed several telemetry drop outs lasting $\approx$ 14 sec every $\approx$ 30--50 sec.}

To acquire a rough estimate of the X-ray flux, for comparison with our other multi-wavelengths measurements, we fit the XMM-Newton X-ray spectrum with a powerlaw $+$ diskbb model, finding an X-ray flux in the $1-10$ keV band of $\sim4.9\times10^{-8}\,{\rm erg\,cm}^{-2}{\rm s}^{-1}$ ($\sim 4$ mJy).
 
   \begin{figure}
\begin{center}
  \includegraphics[width=0.48\textwidth]{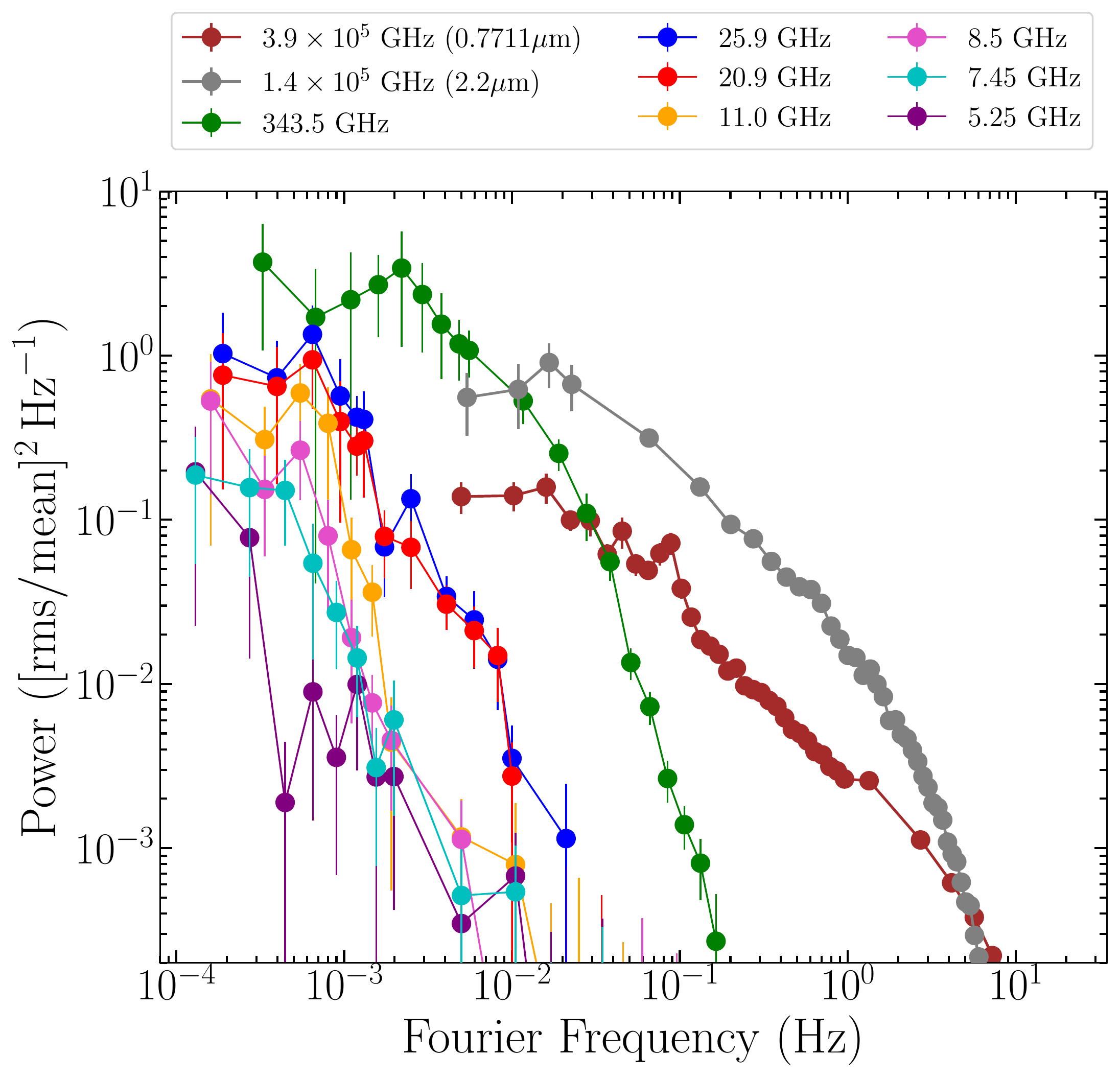}\\
 \caption{\label{fig:psd}Fourier power spectra (PSDs) of optical (0.7711$\mu$m; ULTRACAM), infrared (2.2$\mu$m; HAWK-I), sub-mm (343.5 GHz; ALMA) and radio (5.25 -- 25.9 GHz; VLA) emission from MAXI J1820+070. Note that PSDs of the X-ray bands are shown separately in Figure~\ref{fig:psdX} for clarity. The PSDs shown here were built by stitching together PSD segments created from data imaged/extracted with different time-bin sizes (with the shortest time-scales sampled being 0.01/0.065/2/5 sec for the optical/infrared/sub-mm/radio bands; see \S\ref{sec:psd} for details), in order to circumvent the gaps in the light curves and sample the lower Fourier frequencies. In these PSDs, we observe a clear trend in the shape of the PSDs with electromagnetic frequency band, where the break in the PSDs moves to lower Fourier frequencies as we shift to lower electromagnetic frequency bands. Note that all PSDs shown here have been white-noise subtracted, and the pre-white noise subtracted PSDs are shown in Appendix~\ref{sec:ap_wn}.}
\end{center}
 \end{figure}

\section{Measuring jet timing characteristics}
\label{sec:measure}

\subsection{Light curves}
\label{sec:lc}

Time-resolved multi-band light curves of MAXI J1820+070 are displayed in Figure~\ref{fig:lc} and an average broad-band spectrum is displayed in Figure~\ref{fig:sed}. In the light curves, we observe clear structured variability at all electromagnetic frequencies, in the form of multiple flaring events. Similar flare morphology can be observed between the time-series signals (especially in the radio and sub-mm bands), suggesting that the emission in the different electromagnetic frequency bands is correlated and may show measurable delays. Upon comparing the signals across all of the electromagnetic bands sampled, the variability appears to occur on much faster timescales in the higher electromagnetic bands when compared to the lower electromagnetic bands.
When considering the radio and sub-mm bands, the variability is of higher amplitude in the higher frequency sub-mm band ($\sim 100$ mJy), when compared to the lower frequency radio bands ($\sim5-20$ mJy).  Further, the sub-mm band shows a higher average flux level when compared to the radio bands ($\sim125$ mJy in the sub-mm vs. $\sim46-60$ mJy in the radio bands; see Table~\ref{table:avgfluxes}), indicating an inverted optically thick radio through sub-mm spectrum. The infrared and optical bands appear to not lie on the extension of the radio--sub-mm spectrum, but rather on the steep optically thin portion of the jet spectrum, indicating the jet spectral break lies between the sub-mm and infrared bands (see Figure~\ref{fig:sed}). This result is consistent with previous reports of bright mid-IR emission in excess of the optical emission during the hard state of the outburst \citep{ateldrus1820}.
All of these emission patterns are consistent with the radio, sub-mm, infrared, and optical emission originating in a compact jet, where the higher electromagnetic frequency emission is emitted from a region closer to the black hole (with a smaller cross-section), while the lower electromagnetic frequency emission is emitted from regions further downstream in the jet flow (with larger cross-sections).

  \begin{figure}
\begin{center}
  \includegraphics[width=0.48\textwidth]{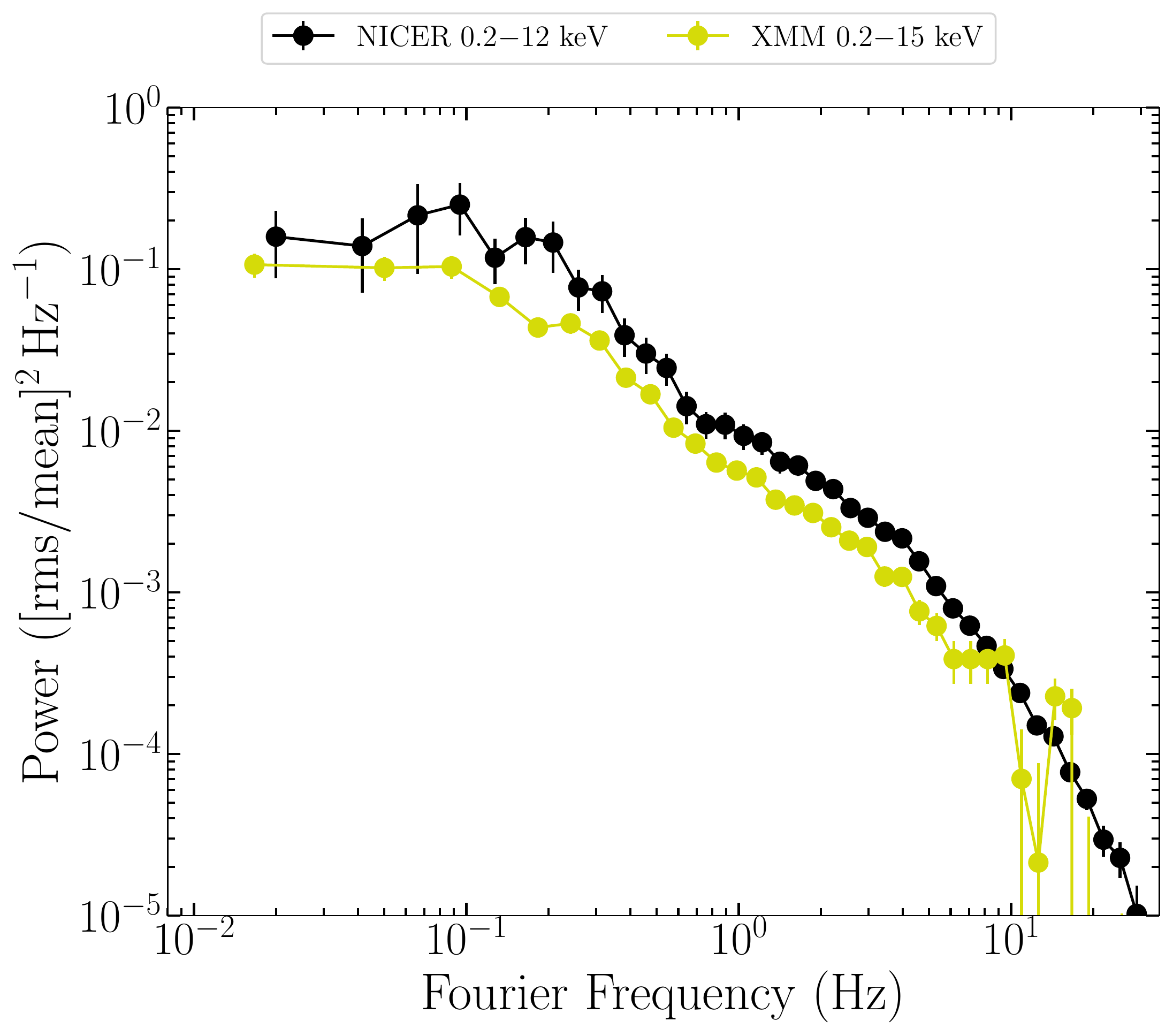}\\
 \caption{\label{fig:psdX}Fourier power spectra (PSDs) of the NICER and XMM-Newton X-ray emission from MAXI J1820+070. Note that the X-ray PSDs shown here have been white-noise subtracted, and the pre-white noise subtracted PSDs are shown in Appendix~\ref{sec:ap_wn}.}
\end{center}
 \end{figure}

\subsection{Fourier Power Spectra}
\label{sec:psd}

To characterize the variability we observe in the light curves of MAXI J1820+070, we opted to perform a Fourier analysis on the data. We use the \textsc{stingray} software package\footnote{\url{https://stingray.readthedocs.io/en/latest/}} for this Fourier analysis \citep{stingray,stingray2}, and Figures~\ref{fig:psd} and \ref{fig:psdX} display the resulting power spectral densities (PSDs).

{As our light curves contain gaps, to build the PSDs over a wide range of Fourier frequencies we stitch together PSD segments created from light curves imaged/extracted with different time-bin sizes. In particular, by building light curves on timescales larger than the gaps, we can manufacture a continuous time-series with which we are able to probe a lower Fourier frequency range.}
For the radio frequency VLA data, we use three PSD segments, built from light curves with 5 sec (final PSD segment is an average over 100 sec chunks), 60 sec (final PSD segment is an average over 15 min chunks), and 240 sec (final PSD segment is an average over 90, 108, 132 min chunks for 20.9/5.9GHz, 8.5/11 GHz, 5.25/7.45 GHz bands, respectively) time-bins. For the sub-mm frequency ALMA data, we use two PSD segments, built from light curves with 2 sec (final PSD segment is an average over 180 sec chunks) and 90 sec (final PSD segment is an average over 50 min chunks) time-bins.
For the infrared/optical frequency data, we use
two PSD segments, built from light curves with 0.0625/0.01 sec (final PSD segment is an average over 15/0.75 sec chunks) and 10/0.5 sec (final PSD segment is an average over 200 sec chunks for both) time-bins.
For the NICER/XMM-Newton X-ray frequency data, we use
only one PSD segment, built from light curves with 0.01/0.004 sec time-bins (final PSD segment is an average over 50/30 sec chunks).
The number of segments/chunk sizes were chosen based on the gap timescales, and to reduce the noise in the PSDs. Further, a geometric re-binning in frequency was applied (factor of $f=0.2$ for radio--sub-mm, $f=0.05$ for infrared/optical, and $f=0.15$ for X-ray, where each bin-size is $1+f$ times larger than the previous bin size) to reduce the scatter at higher Fourier frequencies in all the PSDs. The PSDs are normalized using the fractional rms-squared formalism \citep{bel90}, and white noise has been subtracted\footnote{For the X-ray/optical/IR data, the white noise should be dominated by Poisson/counting noise, while in the radio/sub-mm the white noise is likely due to a combination of atmospheric/instrumental effects.}.
White noise levels were estimated by fitting a constant to the highest Fourier frequencies (see Appendix~\ref{sec:ap_wn}).

The PSDs all appear to display a broken power-law type form, where the highest power occurs at the lowest Fourier frequencies (corresponding to the longest timescales sampled). However, there are clear differences between the PSD shape for the different bands, where the break in the PSDs moves to lower Fourier frequencies as we shift to lower electromagnetic frequency bands. {The same effect can be seen when examining the smallest time-scales (or highest Fourier frequencies) at which significant power is observed in each band (i.e., 10 sec at 343.5 GHz, 100 sec at 20.9/25.9 GHz, and 500 sec at 5.25--11 GHz).} This is the first time an evolving PSD with electromagnetic frequency has been observed from a BHXB. 

\begin{figure}
\begin{center}
  \includegraphics[width=0.45\textwidth]{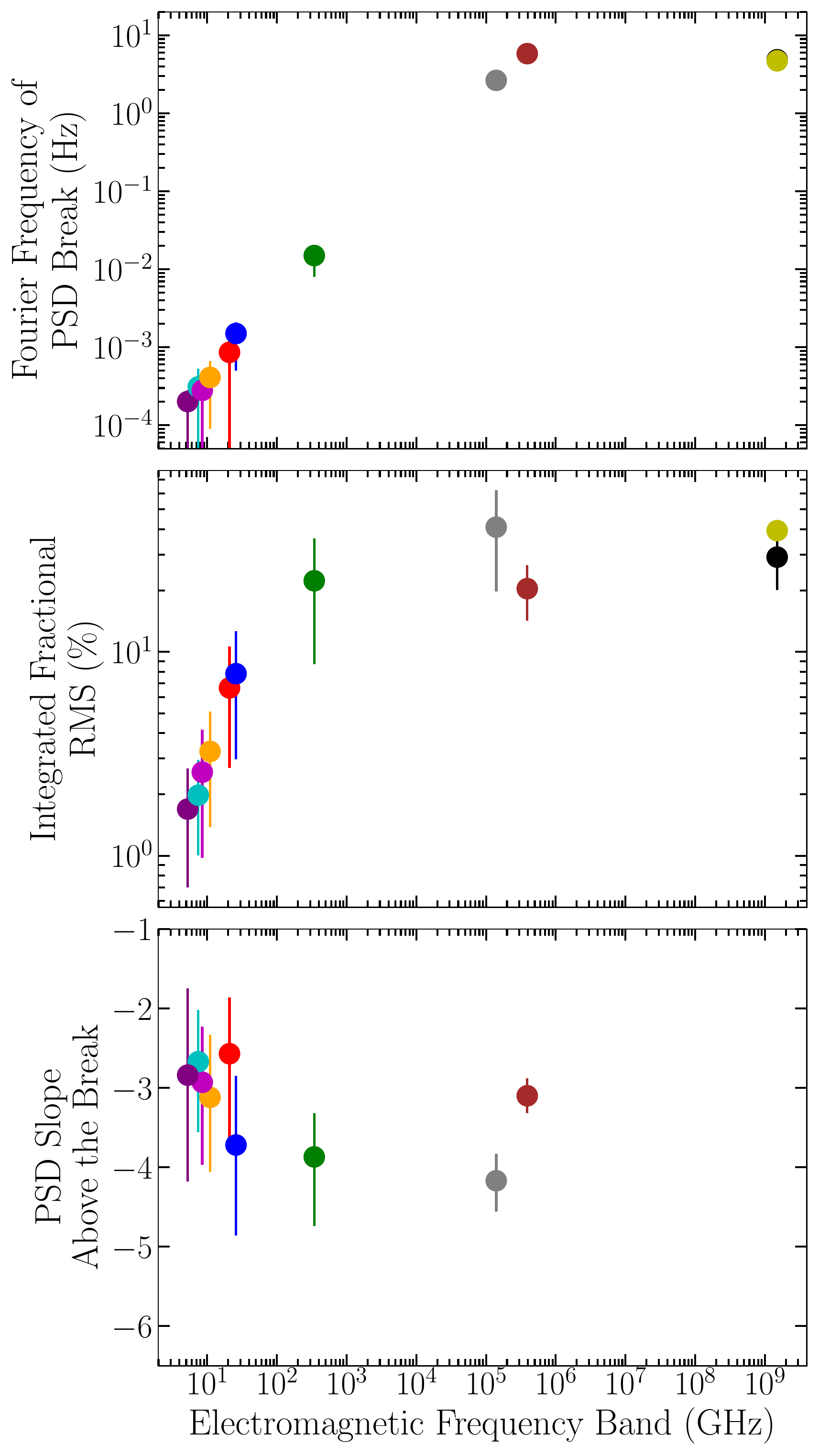}\\
 \caption{\label{fig:psd_metrics}Variability characteristics of the emission from MAXI J1820+070, derived from the PSDs shown in Figure~\ref{fig:psd} \& \ref{fig:psdX}. From \textit{top} to \textit{bottom}, the panels show the Fourier frequency of the PSD break, integrated fractional RMS, and the slope above the PSD break (as we only fit the X-ray PSDs with Lorentzian components, no PSD slope is shown for these bands). The colours of the data points in all panels correspond to the same colours of the electromagnetic frequency bands in Figure~\ref{fig:psd} \& \ref{fig:psdX}. We observe a clear trend with electromagnetic frequency in all of these quantities, except for the slope above the PSD breaks, which remains relatively constant (within error) across the radio--optical bands.}
\end{center}
 \end{figure}

\renewcommand\tabcolsep{10pt}
 \begin{table*}
\caption{PSD modelling results}\quad
\centering
\begin{tabular}{ lcccc}
 \hline\hline
 {\bf Electromagnetic Frequency Band}&{\bf $\bm{f_{\rm break}}$ (Hz) $^\dagger$}& {\bf Slope above break}&{\bf $\bm{z_\nu}$ ($\bm{\times 10^{12}}$ cm)}$^\star$&{\bf $\bm{z_{\rm cross}}$ ($\bm{\times 10^{11}}$ cm)}$^\ddagger$\\[0.15cm]
  \hline
  5.25 GHz&$(2.0^{+2.1}_{-0.8})\times10^{-4}$&$-2.84^{+1.09}_{-1.34}$&$37.6^{+32.7}_{-19.0}$&$2.8^{+2.9}_{-1.5}$\\[0.15cm]
  7.45 GHz&$(3.1^{+2.6}_{-2.2})\times10^{-4}$&$-2.67^{+0.65}_{-0.89}$&$22.2^{+33.4}_{-11.2}$&$1.7^{+2.8}_{-0.9}$\\[0.15cm]
  8.5 GHz&$(2.8^{+3.5}_{-1.4})\times10^{-4}$&$-2.93^{+0.70}_{-1.04}$&$26.7^{+30.1}_{-15.0}$&$2.0^{+2.8}_{-1.2}$\\[0.15cm]
  11.0 GHz&$(4.1^{+3.2}_{-2.6})\times10^{-4}$&$-3.12^{+0.79}_{-0.94}$&$17.8^{+22.9}_{-8.6}$&$1.4^{+1.9}_{-0.7}$\\[0.15cm]
  20.9 GHz&$(8.6^{+4.9}_{-9.9})\times10^{-4}$&$-2.57^{+0.71}_{-1.12}$&$9.1^{+10.6}_{-3.8}$&$0.8^{+1.0}_{-0.4}$\\[0.15cm]
  25.9 GHz&$(1.5^{+0.9}_{-0.6})\times10^{-3}$&$-3.72^{+0.87}_{-1.14}$&$5.2^{+4.2}_{-2.1}$&$0.4^{+0.4}_{-0.2}$\\[0.15cm]
  343.5 GHz&$(1.5^{+0.7}_{-0.5})\times10^{-2}$&$-3.87^{+0.55}_{-0.87}$&$0.5^{+0.3}_{-0.2}$&$0.05^{+0.04}_{-0.02}$\\[0.15cm]
  $2.2\mu$m&$2.7^{+0.4}_{-0.3}$&$-4.18^{+0.34}_{-0.39}$&$(2.9^{+0.8}_{-0.6})\times10^{-3}$&$(2.3^{+1.0}_{-0.8})\times10^{-4}$\\[0.15cm]
  $0.7711\mu$m&$5.8^{+0.8}_{-0.5}$&$-3.10^{+0.22}_{-0.22}$&$(1.3^{+0.3}_{-0.2})\times10^{-3}$&$(1.2^{+0.5}_{-0.4})\times10^{-4}$\\[0.15cm]
  NICER 0.2--12 keV&$4.9^{+0.4}_{-0.4}$&\dots&\dots&\dots\\[0.15cm]
  XMM-Newton 0.2--15 keV &$4.7^{+0.8}_{-0.7}$&\dots&\dots&\dots\\[0.15cm]

 \hline
\end{tabular}\\
\begin{flushleft}
$^\dagger$ {For the NICER/XMM-Newton X-ray PSDs, we take the break frequency to be the ``characteristic frequency" defined in \citealt{belloni02} as $f_{\rm break}=\sqrt{\nu_0^2+\Delta^2}$, where $\nu_0$ is the central frequency and $\Delta$ is the FWHM of the highest Fourier frequency Lorentzian.}\\
$^\star$ Distance downstream from the black hole to the $\tau_\nu=1$ surface. Here we use the formalism, $z_{\nu}=\frac{\beta c \delta}{f_{\rm \,break}}$, and sample from the best-fit $\beta$ distribution along with the known $\theta$ distribution (see \S\ref{sec:psd} and \ref{sec:model} for details).\\
$^\ddagger$ Jet cross-section assuming a conical jet. Here we use the formalism, $z_{\rm cross}=z_{\nu}\tan{\phi}$, and sample from the best-fit $\phi$ distribution (see \S\ref{sec:model} for details).
\end{flushleft}
\label{table:psdfits}
\end{table*}
\renewcommand\tabcolsep{6pt}

To quantitatively characterize the evolving PSDs with electromagnetic frequency band that we observe in our data, we consider two different metrics; (1) integrated fractional rms amplitude {(computed in the Fourier frequency range ${10^{-4}-50}$ Hz)}, and (2) location of the PSD break. Figure~\ref{fig:psd_metrics} displays each of these metrics as a function of electromagnetic frequency band (and also shows the slope after the PSD break as a function of electromagnetic frequency band). To estimate the break frequencies (and slopes after the break) in the PSDs, we have used a Markov Chain Monte Carlo algorithm (MCMC, implemented in the \textsc{emcee} python package; \citealt{for2013}) to fit each PSD with a phenomenological model.

It is commonplace in the BHXB literature to fit X-ray PSDs with Lorentzian components \citep{belloni02}. However, we found that for the radio, sub-mm, infrared, and optical PSDs a Lorentzian component could not fit the highest Fourier frequencies well, as we observe a much steeper damping of the power at these Fourier frequencies when compared to the X-ray PSDs {(although see a discussion in Appendix~\ref{sec:ap_wn} of how windowing and  over-subtraction of white noise can impact the PSDs at higher Fourier frequencies)}.
Therefore, to better model the more severe damping in the power that we see in the radio--optical PSDs, we choose to use a broken power-law component rather than a Lorentzian. Specifically, to fit the radio and sub-mm PSDs we use only a broken power-law component, to fit the infrared/optical PSDs we use a broken power-law component for the highest Fourier frequencies + Lorentzian component(s) for the lower Fourier frequencies, and to fit the X-ray PSDs we use only Lorentzian components. Additionally, we tested how the PSD break varies with the chosen model, finding that different models do not lead to significant differences in the inferred PSD break (see Appendix~\ref{sec:ap_psdfits}).
In our fitting process, we use wide uniform priors for all parameters.
The best fit result is taken as the median of the resulting posterior distributions, and the uncertainties are reported as the range between the median and the 15th percentile (-), and the 85th percentile and the median (+), corresponding approximately to $1 \sigma$ errors.
The best-fit model parameters can be found in Table~\ref{table:psdfits}, while the fits are displayed with residuals in Appendix~\ref{sec:ap_psdfits}.

Figure~\ref{fig:psd_metrics} clearly shows that the PSD break frequency and integrated fractional rms amplitude both change with electromagnetic frequency band, following trends qualitatively consistent with what we might expect from jet model predictions \citep{blandford79,mal14}. 
In particular, the higher variability amplitude observed at the higher electromagnetic frequency jet emitting bands (sub-mm/infrared/optical) reflects the fact that this emission originates from a region with a smaller cross-section, presumably closer to the black hole. The variability amplitudes we measure here in MAXI J1820+070 are consistent with other recent works as well (e.g., mid-IR variability amplitudes of 15--20\% in the BHXB MAXI J1535--571 at $4.85-12.13\mu$m / $2-6\times10^4\, {\rm GHz}$; \citealt{bag18}).
Additionally, the Fourier frequency of the PSD breaks appears to increase with electromagnetic frequency, before leveling off between the optical and X-ray bands.
The factors that set the PSD break frequency for the jet emitting bands (radio--optical) is a complex question. 
In the internal shock model of \citet{mal14}, flux variability is driven by the injection of discrete shells of plasma at the base of the jet with variable speeds. The fastest timescale jet velocity fluctuations are dissipated closest to the base of the jet, as the most rapid variations have led to shells catching up with each other rather quickly. Further downstream in the jet, only the more slowly-varying jet velocity fluctuations still remain to catch up with one another, create shocks, and produce the varying radio emission we observe.  As such, this argues that the PSD break must be related to the distance downstream in the jet (e.g., see Equation 36 of \citet{mal14}, where all the power in a given fluctuation at Fourier frequency, $f$, is dissipated at a distance downstream that scales as $1/f$). In fact, we do find that the PSD breaks scale inversely with electromagnetic frequency band, following the same trend we expect for the size-scale of the jet at different electromagnetic frequencies (i.e., distance downstream from the black hole to the $\tau=1$ surface, $z_{\nu}$; see Table~\ref{table:psdfits}).
However, the situation is likely much more complex, as the \citet{mal14} model also predicts that the distance downstream is connected to the properties of the injected velocity fluctuations (e.g., amplitudes of the fluctuations and power spectral shape of the fluctuations; see Equations 36 and 45 of \citealt{mal14}), and when two shells collide they merge and form a new shell, indicating that different fluctuations can also be correlated with each other. In the following modelling sections of this paper, we will make the simplifying assumption that the PSD breaks are tracing the jet size-scale, but a more detailed investigation of all of the factors that may govern the PSD break will be considered in future work. 

\renewcommand\tabcolsep{10pt}
 \begin{table}
\caption{Radio-radio time lag measurements}\quad
\centering
\begin{tabular}{ cc}
 \hline\hline
 {\bf Frequency Bands }&{\bf Time Lag }\\
 {\bf Compared (GHz)}&{\bf (min)}\\[0.15cm]
  \hline
  25.9/20.9&$1.0^{+0.7}_{-0.9}$\\[0.1cm]
  25.9/11.0&$5.0^{+1.5}_{-1.9}$\\[0.1cm]
  25.9/8.5&$7.0^{+2.6}_{-2.2}$\\[0.1cm]
  25.9/7.45&$8.0^{+2.8}_{-2.5}$\\[0.1cm]
  25.9/5.25$^\dagger$&$>8.0$\\[0.1cm]
  11.0/8.5&$1.0^{+2.6}_{-1.0}$\\[0.1cm]
  11.0/7.45&$2.0^{+2.0}_{-1.4}$\\[0.1cm]
  11.0/5.25$^\star$&$2.0^{+4.5}_{-5.7}$\\[0.1cm]
 \hline
\end{tabular}\\
\begin{flushleft}
$^\dagger$ We do not detect a clear lag between these bands, and thus use the value from the next highest frequency band (7.45 GHz) as an lower limit.\\
$^\star$ The larger errors on this time lag measurement make the measured lag consistent with zero. Note that this CCF is not shown in Figure~\ref{fig:ccfs}.
\end{flushleft}
\label{table:ccfval}
\end{table}
\renewcommand\tabcolsep{6pt}

Unlike the PSD breaks and the variability amplitudes, the slope after the PSD break remains constant across the radio through optical PSDs\footnote{As a test of consistency in our fits, given the near constant slope after the break found in individual fits to the PSDs, we also performed a joint PSD fit. Here we ran the MCMC algorithm, this time tying the slope after the break parameter across the PSDs. This alternate fit yielded PSD breaks that were all consistent (within errors) with the individual fits.}. This constant slope suggests that the mechanism that damps higher Fourier frequency variations in the jet, does not vary with electromagnetic frequency (or distance downstream from the black hole), possibly reflecting the self-similar nature of the jet.

 \begin{figure*}
\begin{center}
  \includegraphics[width=0.45\textwidth]{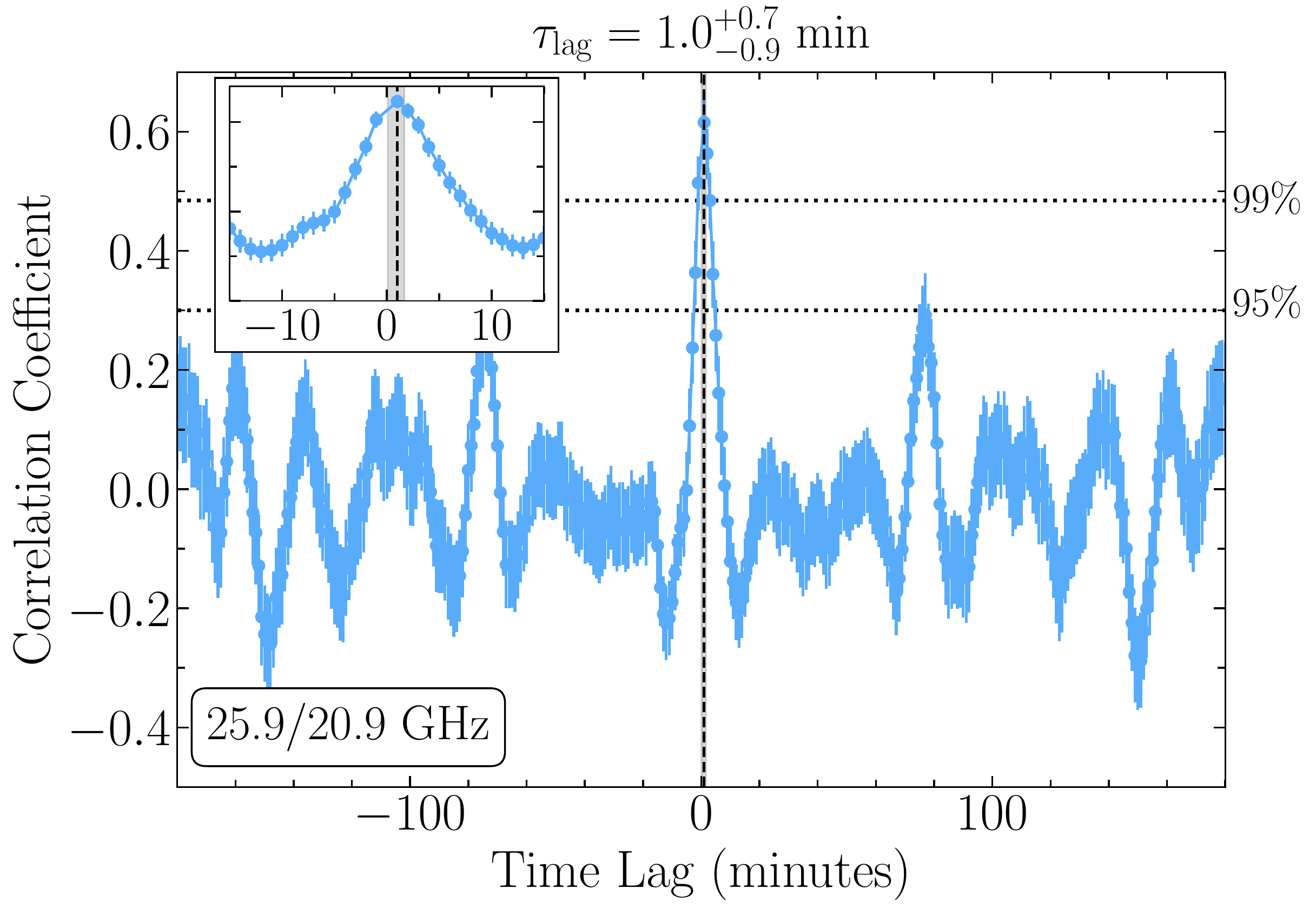}
    \includegraphics[width=0.45\textwidth]{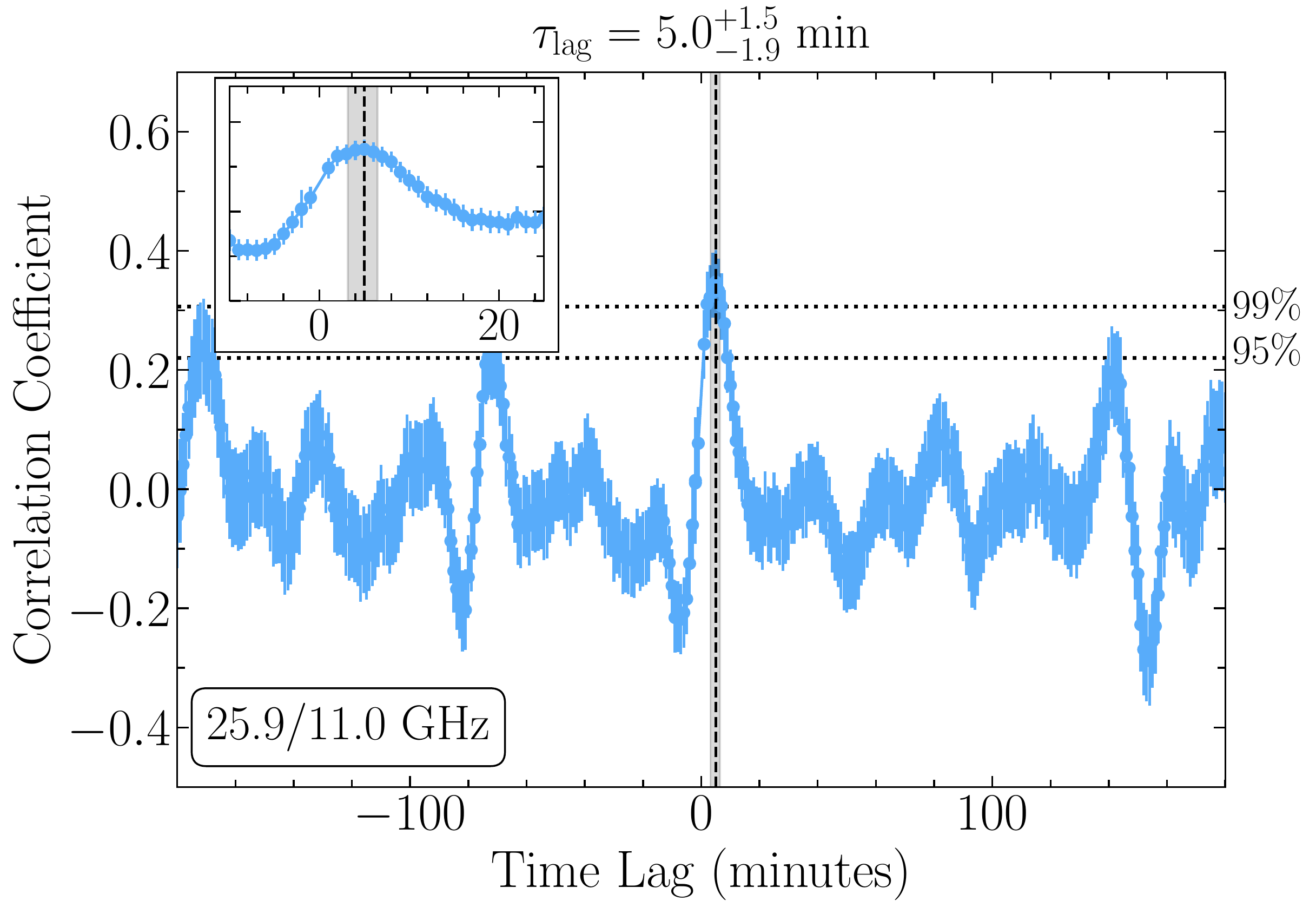}
      \includegraphics[width=0.45\textwidth]{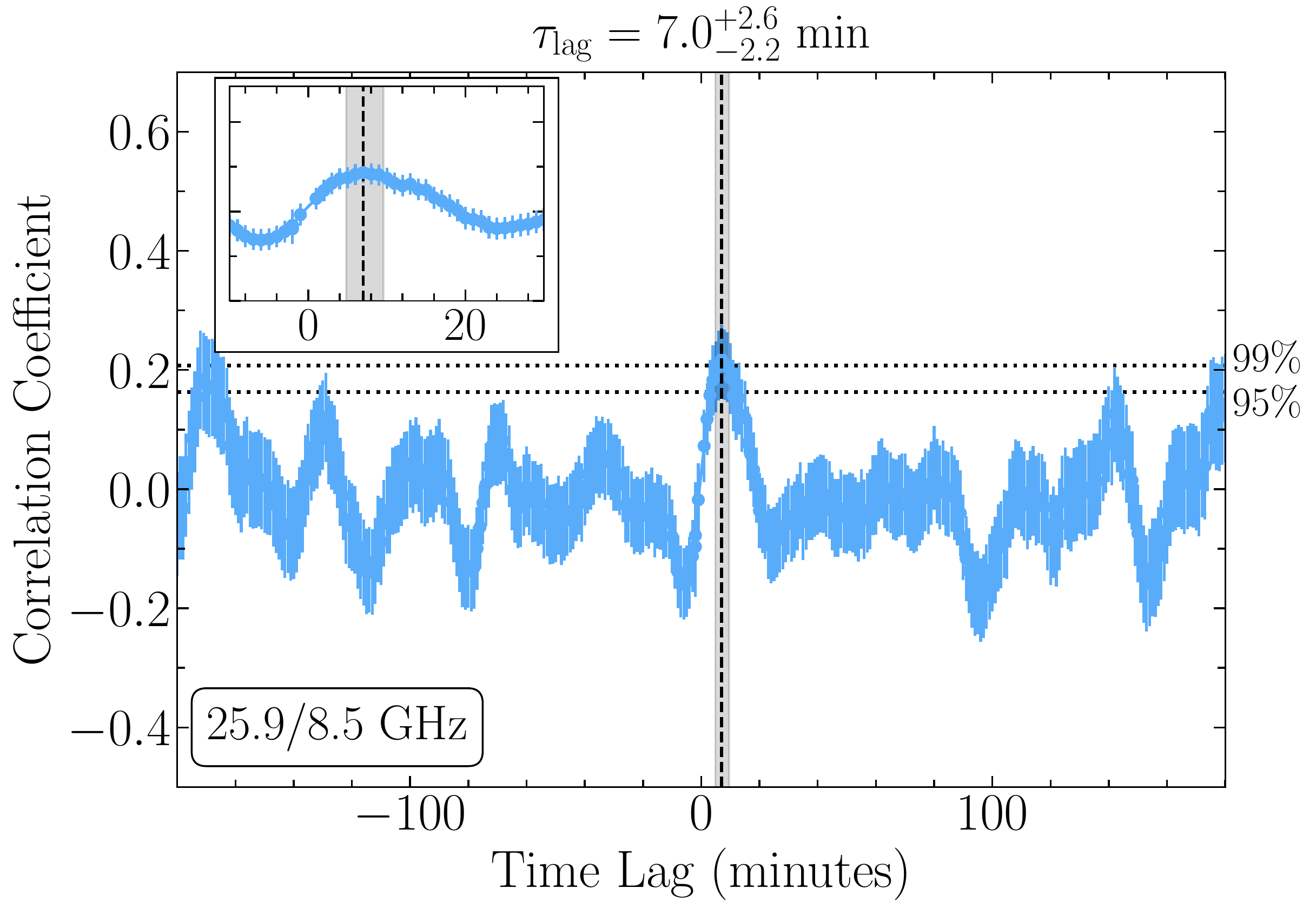}
      \includegraphics[width=0.45\textwidth]{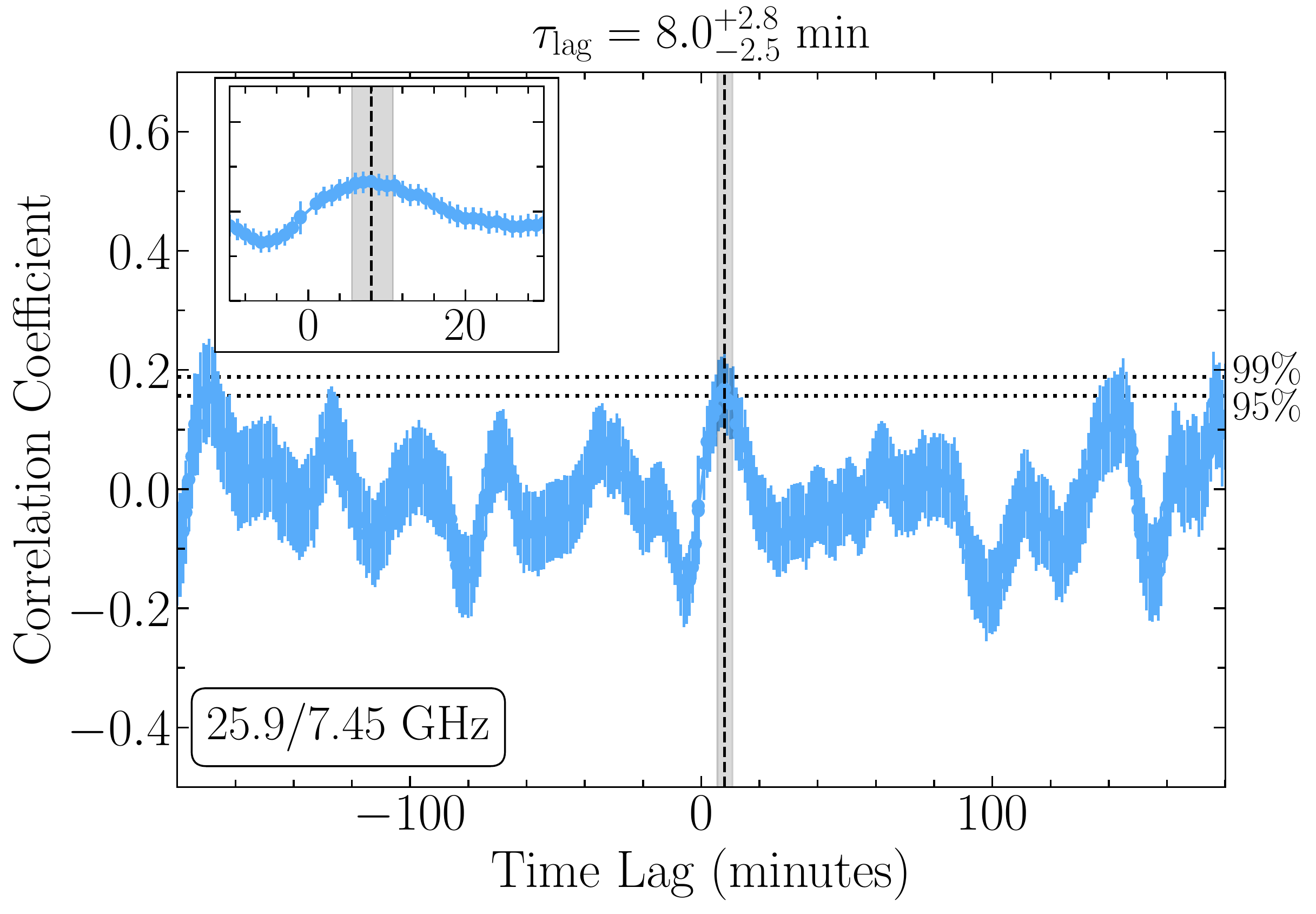}
        \includegraphics[width=0.45\textwidth]{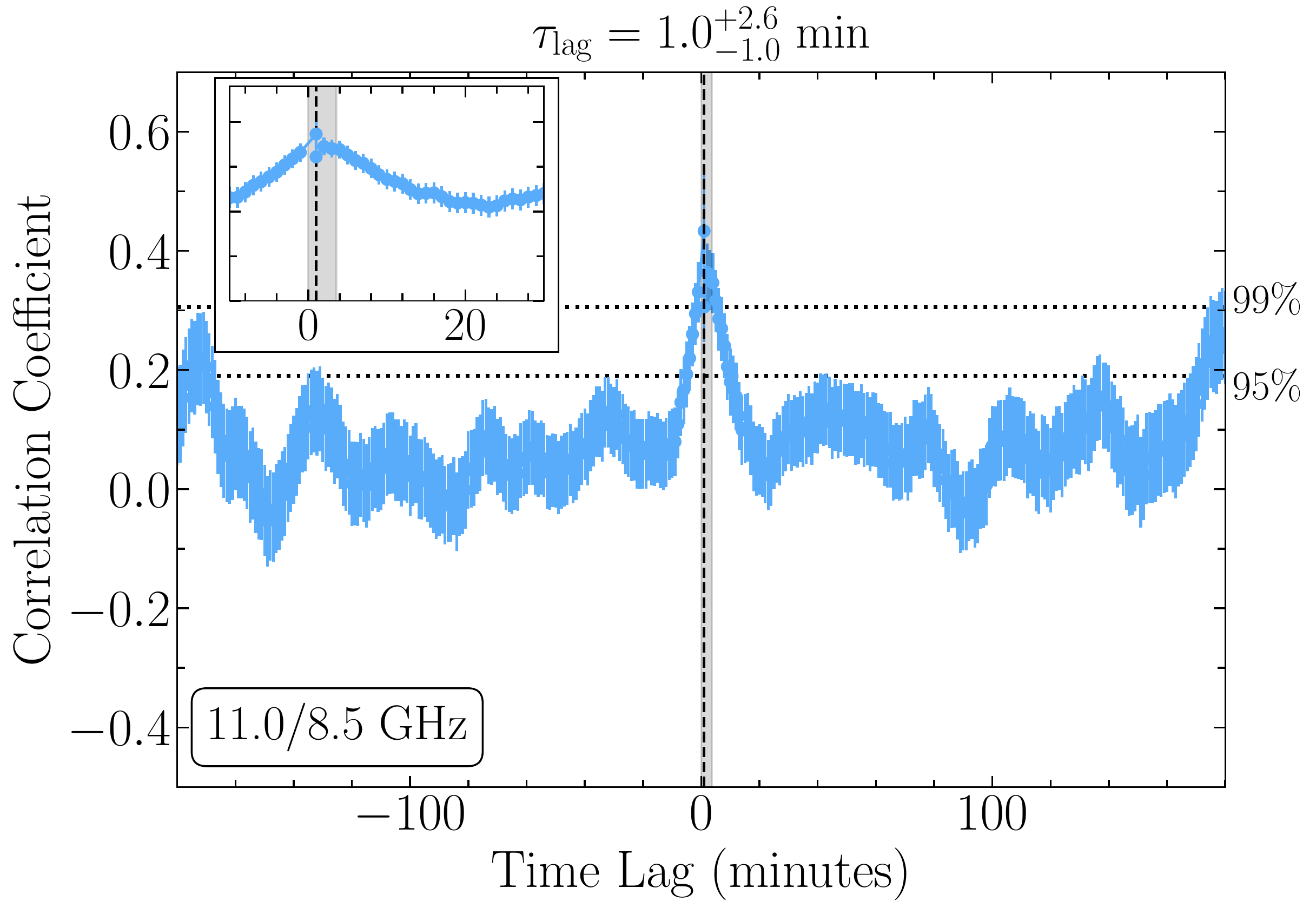}
        \includegraphics[width=0.45\textwidth]{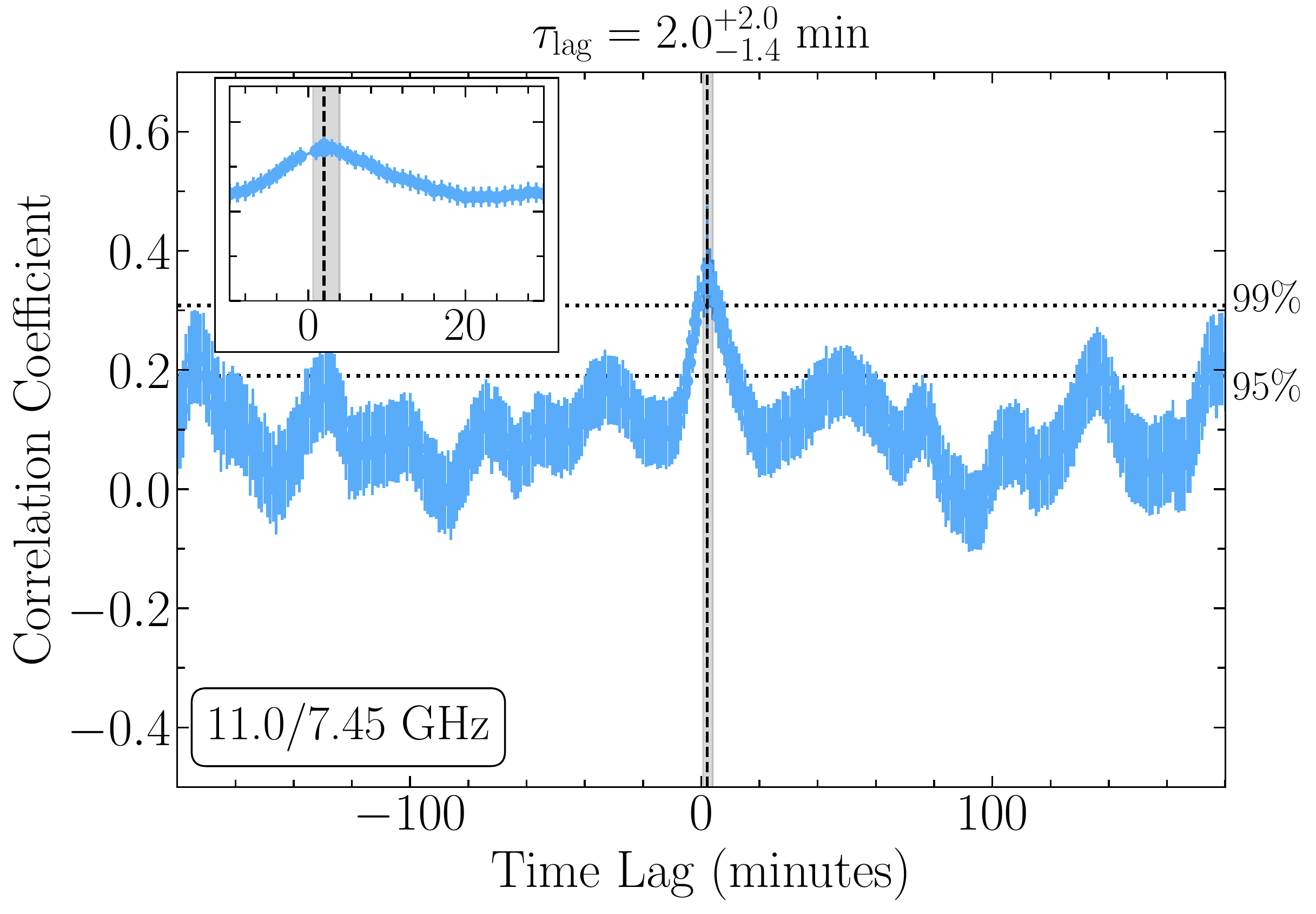}\\
 \caption{\label{fig:ccfs}Cross-correlation functions (CCFs) between emission at different radio frequency bands from MAXI J1820+070 (a positive lag indicates that the lower radio frequency band lags the higher radio frequency band). The panels compare radio signals between 25.9/20.9 GHz (\textit{top left}), 25.9/11.0 GHz (\textit{top right}), 25.9/8.5 GHz (\textit{middle left}), 25.9/7.45 GHz (\textit{middle right}), 11.0/8.5 GHz (\textit{bottom left}), and 11.0/7.45 GHz (\textit{bottom right}). The insets show a zoomed in version of the CCFs near the peak. The black dotted line and gray shading indicates the peak of the CCF and its associated confidence interval (where the measured lag is labelled at the top of each panel). The black dotted lines mark the 95/99\% significance levels (see \S\ref{sec:lag}). We measure clear time lags between the radio signals at various bands on the order of minutes.}
\end{center}
 \end{figure*}
 
  \begin{figure*}
\begin{center}
  \includegraphics[width=0.9\textwidth]{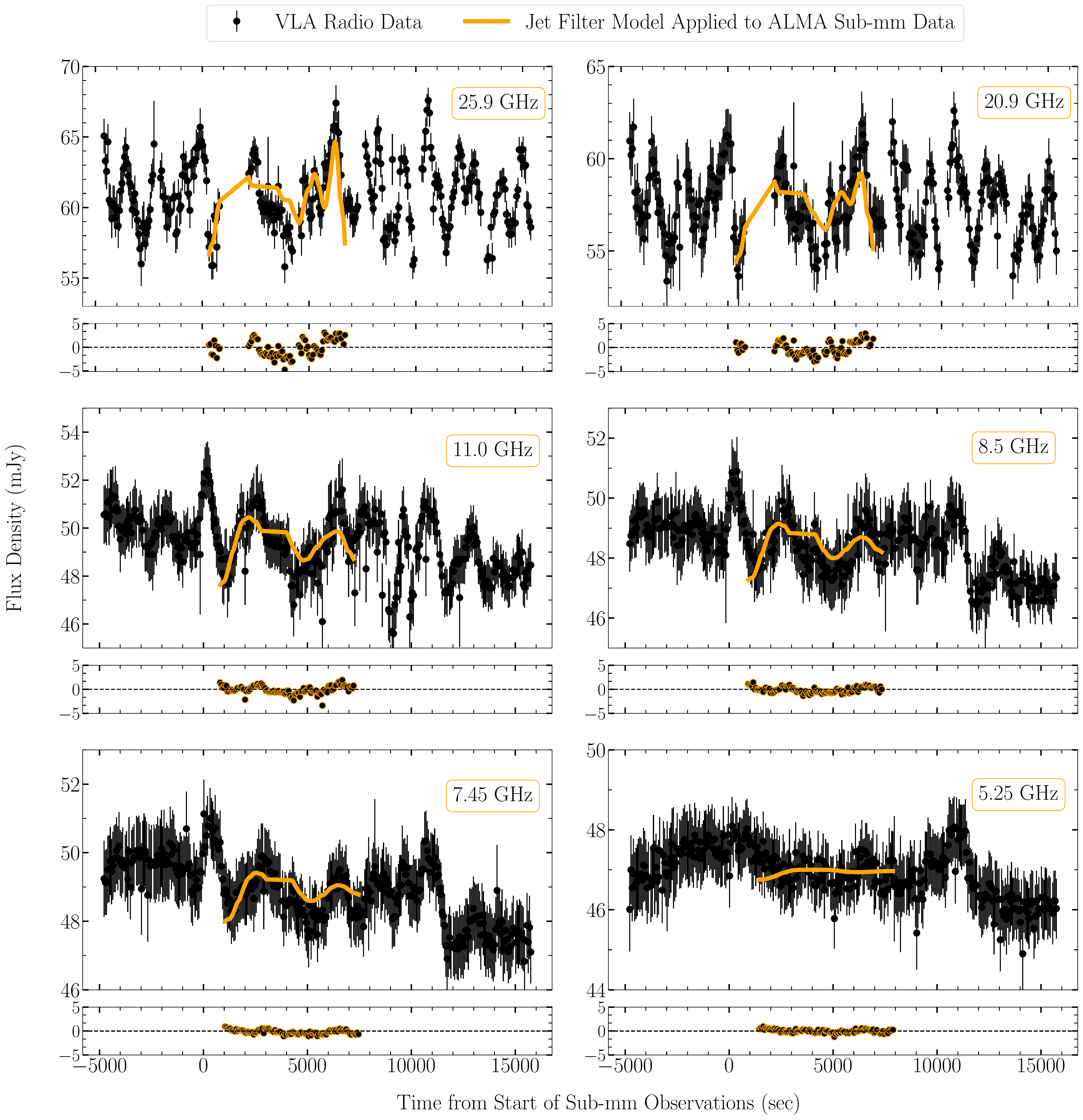}\\
 \caption{\label{fig:almalags_mod}{Modelling the sub-mm--radio lags in MAXI J1820+070. In each panel, the \textit{top} sub-panel displays the VLA radio light curves (black circles), with the result of our jet filter model (solid orange line; best fit parameters in Table~\ref{table:alma_lag}) over-plotted, while the \textit{bottom} sub-panel displays the residuals (data-model/uncertainties). 
 By passing the ALMA sub-mm signal through our jet filter model, we can reproduce the observed radio signals extremely well, and in turn reliably estimate the sub-mm--radio lags in this case.}}
\end{center}
 \end{figure*}

\subsection{Time lags}
\label{sec:lag}
In the light curves of MAXI J1820+070 (Figure~\ref{fig:lc}), we observe similar flaring structures across the different electromagnetic frequency bands, suggesting the time-series signals may be correlated. To test this theory, and search for any time-delays between the different electromagnetic frequency bands, we created cross-correlation functions (CCFs). We follow the same procedure as outlined in \citet{teta19} for our CCF analysis, whereby we use the z-transformed discrete correlation function algorithm (ZDCF; \citealt{alex97,alex13a}) to build the CCFs, the maximum likelihood code of \citet{alex13a} to estimate the peak of each CCF (signifying the strongest positive correlation and best estimate of any time-lag), and perform a set of simulations\footnote{In these simulations, we randomize each light curve (Fourier transform the light curves, randomize the phases of both, then inverse Fourier transform back) to create simulated light curves that share the same power spectra as the real light curves, and then calculate the CCF for each randomized case. Significance levels are then based on the fraction of simulated CCF data points (at any lag) above a certain level. See \S3.2 of \citet{teta19} for more details.} allowing us to quantify the probability of false detections in the CCFs. We note that the ZDCF method uses a different binning criterion when compared to the classic discrete correlation function of \citealt{ed88}. In particular, equal population binning is used, where the bins are not equal in time-lag width.

  \begin{figure*}
\begin{center}
\includegraphics[width=0.45\textwidth]{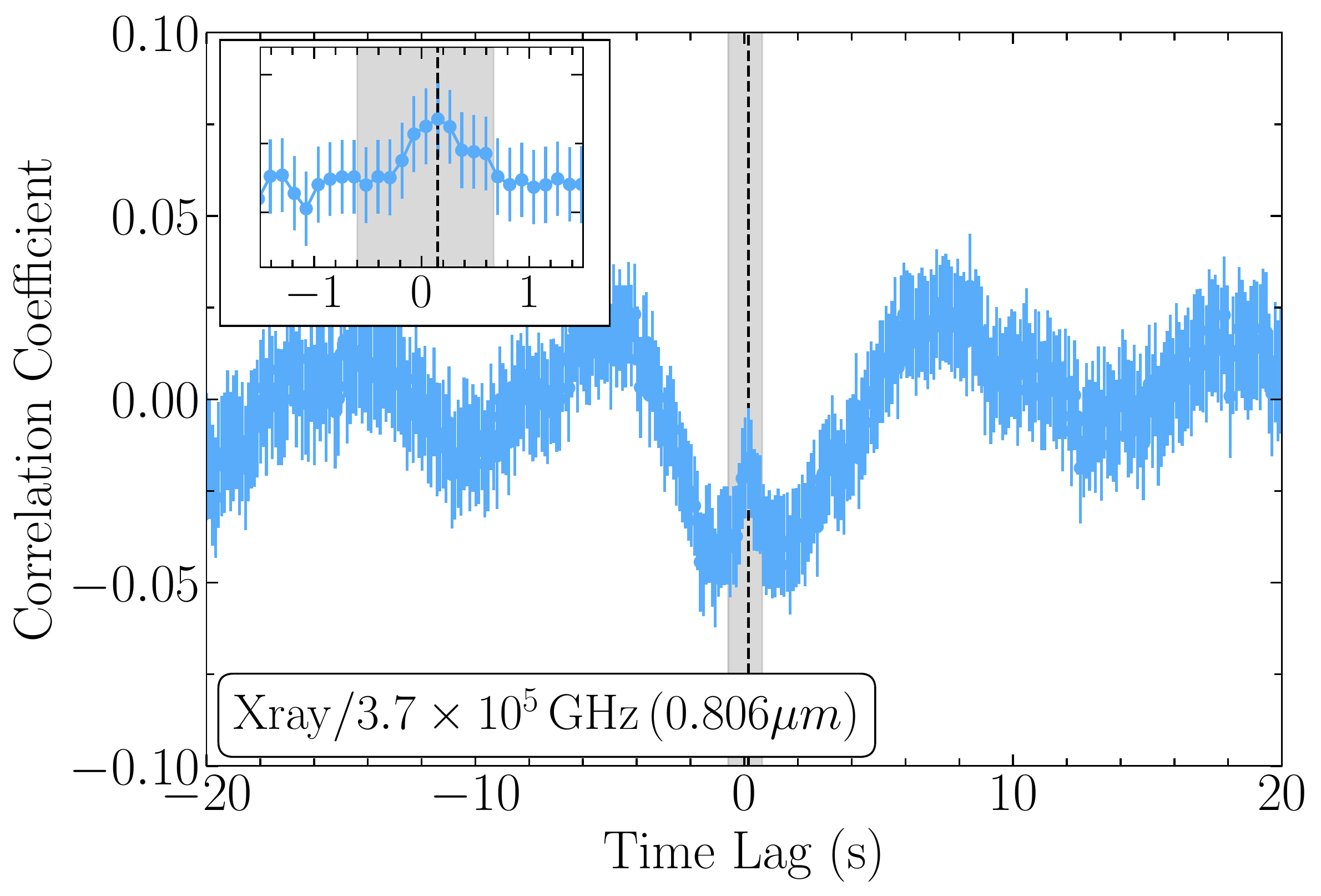}
  \quad\quad
  \includegraphics[width=0.45\textwidth]{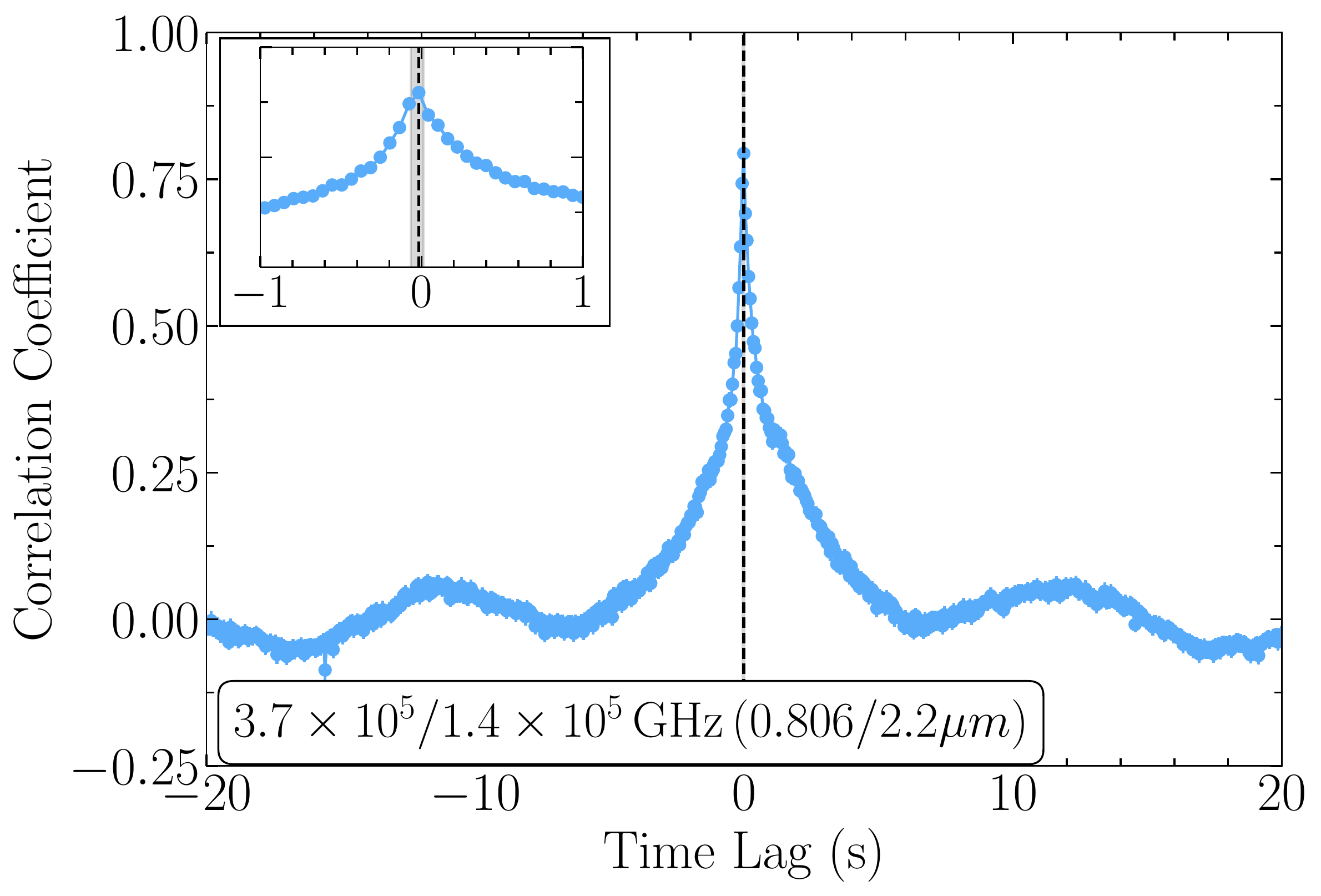}
  \\
 \caption{\label{fig:ccfs_oirx}Cross-correlation functions (CCFs) between XMM-Newton X-ray/optical (\textit{left}) and optical/infrared (\textit{right}) emission from MAXI J1820+070 (a positive lag indicates that the lower electromagnetic frequency band lags the higher electromagnetic frequency band). The insets show a zoomed in version of the CCFs near the peak, while the black dotted line and gray shading indicates the peak of the CCF and its associated confidence interval. The optical/infrared CCF shows a lag consistent with zero ($-18^{+30}_{-50}$ ms) , while the X-ray/optical emission shows a much more complex CCF structure {(similar to that reported by \citet{pai19} for other observations of this source)}, including a peak at $151^{+500}_{-700}$ ms.}
\end{center}
 \end{figure*}

\renewcommand\tabcolsep{6pt}
\subsubsection{Radio -- radio lags}
\label{sec:rrlag}

Figure ~\ref{fig:ccfs} displays the CCFs comparing the time-series signals between the different radio frequency bands sampled. We measure clear time-lags between several radio bands on timescales between 1 -- 8 min (see Table~\ref{table:ccfval}), indicating that the radio signals are correlated. The lower radio frequency bands always lag the higher radio frequency bands, and this lag increases as the observing frequency decreases in the comparison band. As lower radio frequencies probe further downstream from the black hole in the jet flow, these patterns in radio--radio time-lag are consistent with the measured lags tracing the propagation of material along the radio emission regions in the jet.
{We note that all of our radio-radio CCFs display symmetric peaks at the measured time-lag, which reach or exceed the 99\% significance level, and also show roughly the same width, indicating that the observed lags likely correspond to variations on comparable timescales. Therefore, we consider the detected time lags statistically significant, and are confident they are tracking a real correlation between the light curves. }
However, we also note that in all our CCFs there exist secondary peaks (which are likely the result of red noise) at $\sim$-170/140 min, that can at times approach the 95/99\% significance levels, but still remain less significant than the measured lags.
As the time-lags at these secondary peaks remain constant (within error) across the radio bands, and the characteristic radio flaring timescale is a few hundred seconds (as seen in Figure~\ref{fig:lc}), we believe these secondary peaks are the result of the CCF algorithm matching the largest flares in each band to adjacent flares in the time-series, and thus do not represent physical lags.

\subsubsection{Sub-mm -- radio lags}
\label{sec:srlag}

When running the ZDCF algorithm to compare the ALMA sub-mm emission and the VLA radio emission, we did not detect any measurable lags. While we do tend to observe broad peaks in these CCFs, they do not reach the 95/99\% significance levels. This likely indicates that the lags are identified by the CCF algorithm, but they are not statistically significant. Upon examining the sub-mm and radio light curves in Figure~\ref{fig:lc}, it is clear that the sub-mm signal is flaring on much more rapid timescales than the radio flaring. Therefore, we believe that the reason we can not significantly detect a lag between the sub-mm and radio bands in the CCFs is that the frequency separation between the sub-mm and radio bands is too large, and the sub-mm signal has been significantly smoothed out by the time it reaches the radio bands.

As the CCF method does not appear to be sufficient to measure sub-mm -- radio lags in this case, we designed an alternative method in which we essentially pass the sub-mm signal through a jet filter, and compare the resulting signal to our radio signals. To apply this jet filter, we delay and smooth the sub-mm signal with a Gaussian kernel (with smoothing time-scale, $\sigma_{\rm smooth}$). To solve for the delays and smoothing timescales (both of which will vary with the comparison radio band), we simultaneously fit our model to all radio frequencies (5.25--25.9 GHz) using an MCMC algorithm. In this fitting process, we use wide uniform priors for all parameters. As we wish to obtain a jet model independent estimate of the sub-mm-radio lags (mimicking the CCF algorithm), we do not tie the delays or smoothing timescales between radio bands together in the fits (in a \citet{blandford79} jet model, we would expect both delay and smoothing timescale to inversely scale with radio frequency).
Further, we only use the ALMA sub-mm signal from MJD 58220.365 (08:45 UT) onward in our modelling, as we find that the ALMA observations do not start early enough to sample all of the sub-mm emission that, when smoothed and delayed, will contribute to the large radio flare at $\sim$08:55.

The best-fit jet filter parameters and lags are displayed in Figure~\ref{fig:almalags_mod} and Table~\ref{table:alma_lag}.
The best fit result is taken as the median of the resulting posterior distributions, and the uncertainties are reported as the range between the median and the 15th percentile (-), and the 85th percentile and the median (+), corresponding approximately to $1 \sigma$ errors.
Our jet filter method is able to reproduce the radio frequency signals extremely well, and the resulting sub-mm-radio lags follow the same pattern as the radio-radio lags (where lower radio frequencies lag the higher sub-mm frequency, and the lag increases as the radio frequency decreases in the comparison band).
Therefore, we are confident that the modelled sub-mm-radio lags are tracking a real correlation between the sub-mm and radio light curves, and in turn represent the propagation of material along the jet flow. The best-fit smoothing timescales also scale inversely with radio frequency, and the characteristic length scales ($z_{\rm smooth}$ in Table~\ref{table:alma_lag}) implied by these smoothing timescales, are consistent with our estimates for the jet cross-sections (based on our PSD measurements; $z_{\rm cross}$ in Table~\ref{table:psdfits}), further reinforcing the validity of our jet filter method.

\renewcommand\tabcolsep{8pt}
 \begin{table}
\caption{Sub-mm-radio lag modelling results}\quad
\centering
\begin{tabular}{ cccc}
 \hline\hline
 {\bf Frequency Bands }&{\bf Time Lag }&{\bf $\bm{\sigma_{\rm smooth}}$}&{\bf $\bm{z_{{\rm smooth}}}$}\\
 {\bf Compared (GHz)}&{\bf (min)}&{\bf (sec)}&{\bf ($\bm{\times 10^{11}}$ cm)}$^\dagger$\\[0.15cm]
  \hline
  343.5/25.9&$4.5^{+0.2}_{-0.4}$&$6.5^{+0.6}_{-0.6}$&$0.5^{+0.1}_{-0.1}$\\[0.1cm]
  343.5/20.9&$5.5^{+1.2}_{-1.0}$&$9.2^{+1.6}_{-1.2}$&$0.7^{+0.2}_{-0.2}$\\[0.1cm]
  343.5/11.0&$12.6^{+1.2}_{-1.3}$&$13.8^{+2.1}_{-1.6}$&$1.1^{+0.3}_{-0.2}$\\[0.1cm]
  343.5/8.5&$15.7^{+1.7}_{-1.8}$&$17.7^{+3.1}_{-2.1}$&$1.4^{+0.4}_{-0.3}$\\[0.1cm]
  343.5/7.45&$16.7^{+2.2}_{-1.7}$&$21.8^{+3.8}_{-2.4}$&$1.8^{+0.5}_{-0.4}$\\[0.1cm]
  343.5/5.25&$23.6^{+4.5}_{-3.1}$&$37.5^{+8.6}_{-4.2}$&$3.1^{+1.0}_{-0.7}$\\[0.1cm]

 \hline
\end{tabular}\\
\begin{flushleft}
$^\dagger$ Computed using the formalism, $z_{{\rm smooth}}=\beta_{\rm exp} c \delta \sigma_{\rm smooth}$. Here we sample from the best-fit $\Gamma$ and $\phi$ distributions along with the known $i$ distribution (see \S\ref{sec:srlag} and \ref{sec:model} for details). The expansion velocity is computed using the relation,  $\beta_{\rm exp}=\tan\phi\left[\Gamma^2\{1-(\beta\cos i)^2\}-1\right]^{0.5}$ \citep{tetarenkoa17}.
\end{flushleft}
\label{table:alma_lag}
\end{table}

\subsubsection{X-ray--optical--infrared lags}
\label{sec:xoilag}
Figure~\ref{fig:ccfs_oirx} displays the CCFs comparing the time-series signals between the X-ray, optical, and infrared bands. {The X-ray/optical CCF displays a complex structure, including a peak at $\sim 150$ ms, an anti-correlation dip between $\sim-2$ and $4$ sec, and broad peaks at larger lags.} These CCF features are similar to those reported in \cite{pai19} for observations taken 5 days after those reported in this paper. An X-ray/optical lag on the order of hundreds of ms has been seen in several BHXBs already (e.g., \citealt{gan17}), and is often interpreted as tracing the propagation of accreted material from the X-ray emitting regions in the accretion flow to the optical emitting regions in the jet base. {The asymmetric anti-correlation dip could be a result of superposition of a symmetric anti-correlation with positive timing humps, due to reprocessing or QPOs \citep{vel11}.}
The optical/infrared emission also appears to be highly correlated, displaying a strong CCF peak that indicates a lag consistent with zero ($-18^{+30}_{-50}$ ms). These results, combined with the broad-band spectrum shown in Figure~\ref{fig:sed}, suggest that the infrared and optical emission may both originate in the optically thin innermost jet base region (although the jet may not be the only contributor to the optical emission from the system; \citealt{vel19,kos20}), and that any lag between the optical/infrared bands may be attributed to the synchrotron cooling time of the electrons in the jet. In this case, we can use the optical/infrared lag to estimate the magnetic field strength in the jet base region ($B$). The synchrotron cooling time is defined as,

\begin{equation}
    t_{\rm sync}=\frac{6\pi m_e c}{\sigma_T B^2 \gamma}=7.8\left(\frac{10^4 {\rm G}}{B}\right)\left(\frac{1}{\gamma}\right)\,{\rm s}
\end{equation}

Here the Lorentz factor of the electrons, $\gamma=(9.1\times10^{-4})(\nu/B)^{1/2}$, where $\nu$ is in units of Hz and $B$ is in units of Gauss. In turn, the synchrotron cooling induced lag can then be written as,

\begin{equation}
    \tau_{\rm opt/IR}=(8.6\times10^{11})B^{-3/2}(\nu_{\rm opt}^{-1/2}-\nu_{\rm IR}^{-1/2})\,{\rm s}
\end{equation}

Substituting in $\tau_{\rm opt/IR}<68$ ms, yields a magnetic field strength constraint of $B>6\times10^3$ G in the jet base region. This constraint is consistent with magnetic field strength estimates made using the measured spectral break frequency in the broad-band spectrum of the compact jet launched by several different BHXBs in the hard accretion state ($B\sim10^4$ G; e.g., \citealt{chat,gan11,rus14,rusluctet20}).

\section{Modelling jet timing characteristics}
\label{sec:model}

In \S\ref{sec:measure}, we presented measurements of several quantities characterising the variable emission we observed from MAXI J1820+070 (e.g., PSD breaks, lags). All of these measured quantities can be predicted by the compact jet model of \cite{blandford79}, depending on jet properties: jet power ($P$), speed ($\beta=v/c$), opening angle ($\phi$), inclination angle of the jet axis ($i$), distance ($D$), and particle properties (filling factor $f$, equipartition fraction\footnote{In this model, the particle pressure is a fixed fraction of the magnetic pressure, such that, $p_{\rm part}=\frac{p_{\rm mag}}{\xi_B}$.} between the particles and the magnetic field $\xi_B$).

Following the formalism outlined in \cite{heinz06}, the observed jet flux density (in units of  ${\rm erg\,s}^{-1}{\rm cm}^{-2}{\rm Hz}^{-1}$) at electromagnetic frequency, $\nu$, can be expressed as,
{
\begin{equation}
F_\nu=\frac{5.1}{4\pi D^2}z_\nu^{17/8} \sin{i}^{7/8} \phi^{9/8} C_0 C_1^{-7/8} \delta^{1/4} \left(\frac{\nu}{\nu_{\rm ref}}\right)^\alpha  
\label{eq:Fnu}
\end{equation}}
where $z_\nu$ represents the distance downstream (along the spine of the jet, in units of cm) from the black hole to the $\tau_\nu(z_\nu)=1$ surface, $\delta=\Gamma[1-\beta\cos{i}]^{-1}$ represents the Doppler factor, $\Gamma=[1-\beta^2]^{-0.5}$ represents the bulk Lorentz factor, $\alpha$ represents the spectral index for the time-averaged spectrum\footnote{The \citet{blandford79} jet model results in a flat spectrum ($\alpha=0$), but the radio--sub-mm MAXI J1820+070 spectrum is clearly inverted, so we add an additional term to account for this here.}, $\nu_{\rm ref}$ represents a reference frequency to anchor the time-averaged spectrum (we set $\nu_{\rm ref}=20.9$ GHz, in the middle of the electromagnetic frequency range covered), and $C_0$/$C_1$ are the constants \citep{ryblig},

\begin{eqnarray}\nonumber
C_0&=&(2.4\times10^{-17})\left(\frac{2 \xi_B^{3/4} f}{1+\xi_p}\right)\left(\frac{2}{1+\xi_B}\right)^{7/4} \left(\frac{8.4 {\rm GHz}}{\nu}\right)^{1/2}\\
&\equiv&(8.4)^{1/2}X_0\nu^{-1/2} \,\,\,\, {\rm erg\,s}^{-1}{\rm cm}^{-3}{\rm Hz}^{-1} \\  \nonumber
C_1&=&(2.3\times10^{-12})\left(\frac{2 \xi_B}{1+\xi_p}\right)\left(\frac{2}{1+\xi_B}\right)^{2} \left(\frac{8.4 {\rm GHz}}{\nu}\right)^{3} f\\ \nonumber
&\equiv&(8.4)^{3}X_1\nu^{-3}\,\,\,\, {\rm cm}^{-1}  \nonumber
\end{eqnarray}
{Here $\xi_B$ represents the equipartition fraction, we set the proton contribution term $\xi_p=0$ to signify a purely leptonic jet with no protons\footnote{$\xi_p$ is defined in terms of the proton contribution to the particle pressure; $p_{\rm proton}=\frac{\xi_p}{1+\xi_p}p_{\rm part}$.}, and we group constant terms (including $\xi_p$ and $\xi_B$) into the $X_0$ and $X_1$ constants}.

The kinetic jet power (including the counter-jet contribution, but not the kinetic energy from the bulk motion) can be expressed {as a function of distance downstream in the jet ($z$),} as,
\begin{equation}
W=2\left[4 p \Gamma^2 \beta c \pi (\phi z)^2\right]\,\,\,\, {\rm erg\,s}^{-1}
\label{eq:W}
\end{equation}
where jet pressure $p=\sqrt{\frac{\sin{i}}{2 C_1 \delta^2 \phi z_\nu}}\left(\frac{z}{z_\nu}\right)^{-2}$, and $c$ is the speed of light.
The power contribution from the kinetic energy due to the bulk motion can be expressed as, $W_{\rm KE}=[\Gamma-1]W$. Therefore, the total power becomes $P=W + W_{\rm KE}=W\Gamma$.

 \begin{figure*}
\begin{center}
  \includegraphics[width=1\textwidth]{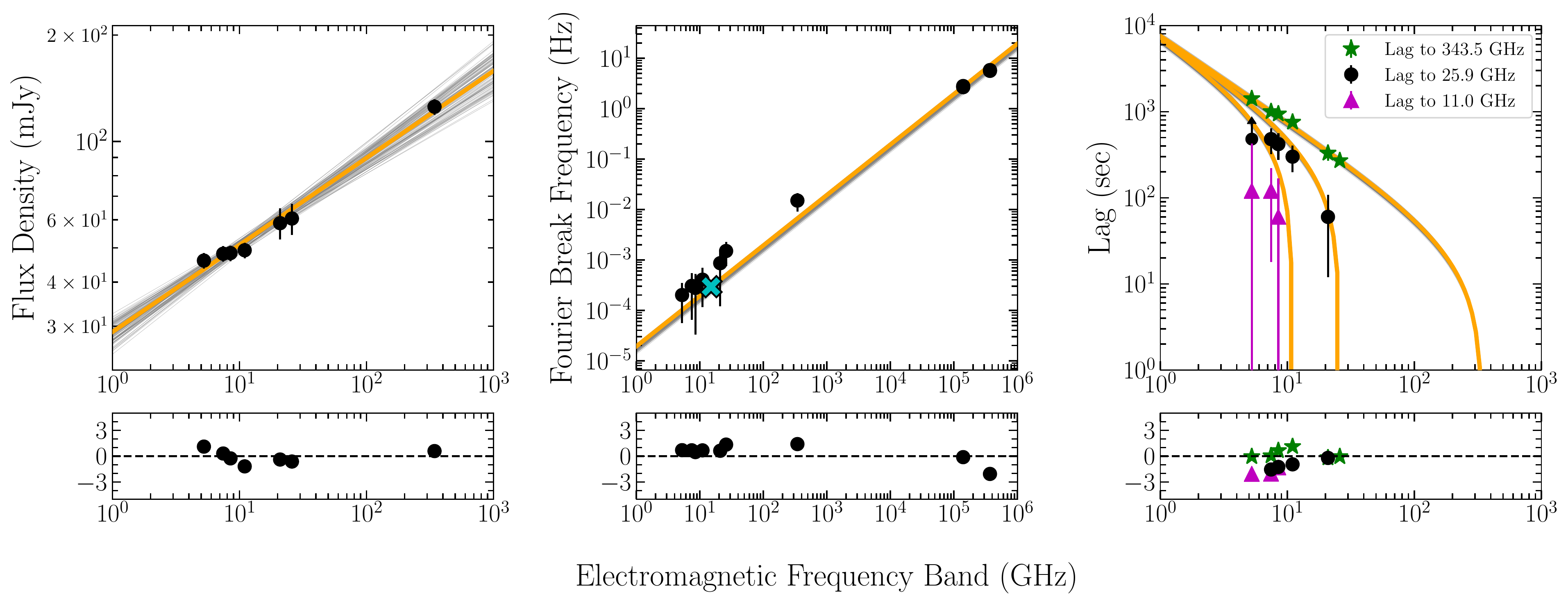}\\
 \caption{\label{fig:jetmod} Modelling the jet timing properties in MAXI J1820+070. The panels from \textit{left} to \textit{right} display the time-averaged jet spectrum, the Fourier break frequency, and the time-lags to 343.5, 25.9, and 11 GHz, all as a function of electromagnetic frequency band. In each panel, the \textit{top} sub-panel displays the data (black circles, green stars, and magenta triangles) with the result of the best-fit jet model over-plotted (solid orange line represents the model, where the thin gray lines show the final positions of all the walkers in the MCMC run, to represent the $1\sigma$ confidence interval; see Table~\ref{table:jetmodel}), while the \textit{bottom} sub-panel displays the residuals (data-model/uncertainties). The cyan X in the middle panel represents an independent measurement of the jet size scale at 15 GHz from {VLBA imaging}, projected onto this plane (i.e., $\beta c \delta/z_{\rm vlba}$; see \S\ref{sec:model}). 
 {We do not include infrared/optical fluxes in the modelling as a rigorous absolute flux calibration was not performed on this data, and we do not include X-ray/optical/infrared lags in the modelling as our model only describes the partially self-absorbed optically thick portions of the jet.}
 The \citet{blandford79} jet model can reproduce these three data dimensions reasonably well. }
\end{center}
 \end{figure*}

 \begin{figure*}
\begin{center}
  \includegraphics[width=0.85\textwidth]{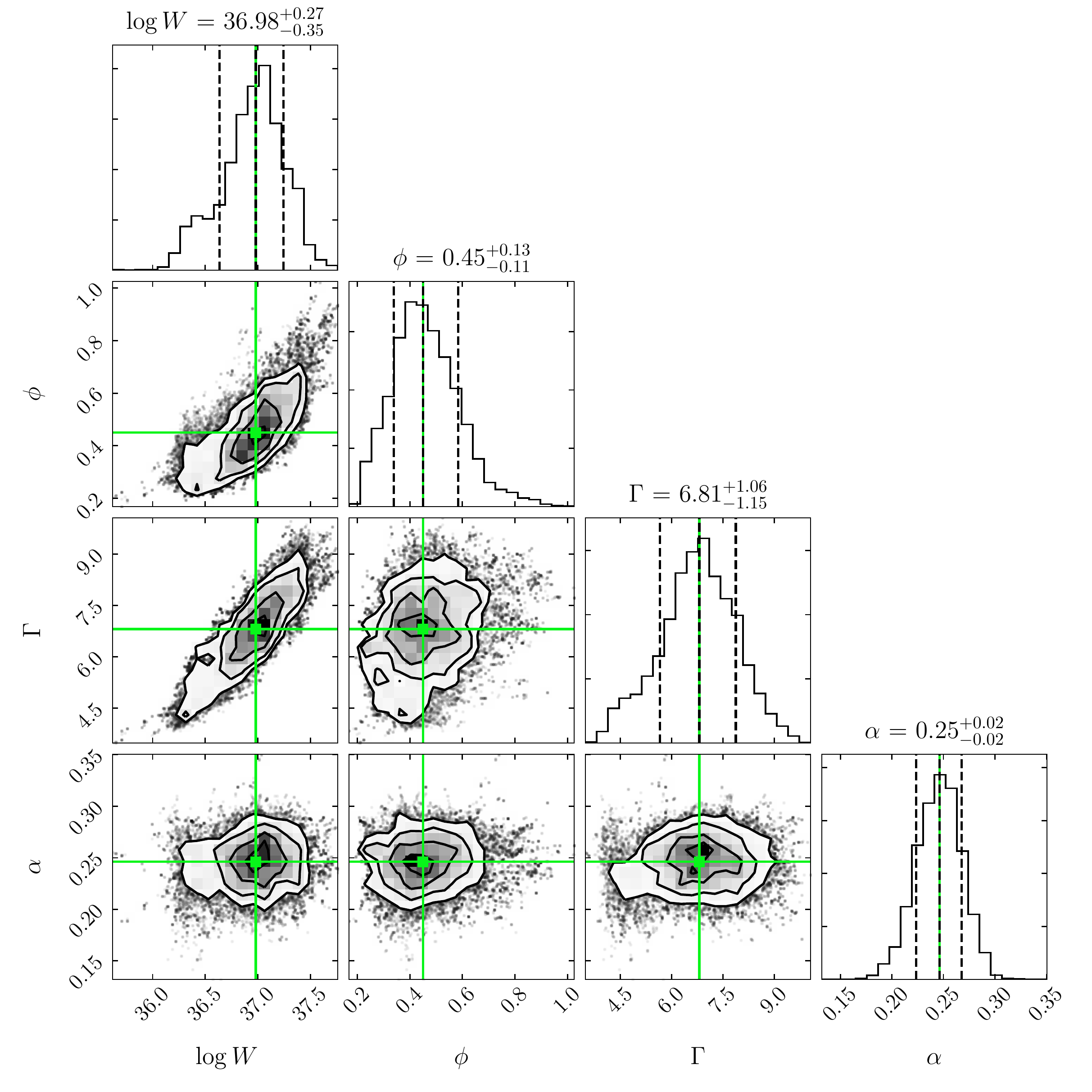}\\
 \caption{\label{fig:corner} Corner plots displaying the results of the MCMC \citet{blandford79} jet model fit (see Table~\ref{table:jetmodel}). The panels show the histograms of the one dimensional posterior distributions for the free model parameters, and the 2-parameter correlations, with the best-fit values of the free parameters indicated by green lines/squares. These corner plots were created with the \textsc{corner} plotting package \citep{cornerref}.}
\end{center}
 \end{figure*} 
 
Rearranging Equation~\ref{eq:W}, and substituting in the expression for jet pressure, yields,

\begin{equation}
\label{eq:z}
z_\nu=\left(\frac{W}{8 \Gamma^2 \beta c \pi \phi^{3/2}}\right)^{2/3}\left(\frac{2 X_1 (8.4)^3 \delta^2}{\sin{i}}\right)^{1/3} \nu^{-1}\,\,\,\, {\rm cm}
\end{equation}
Note that in \S\ref{sec:measure}, we have discussed the factors that may govern the PSD break. While we realized that this was far from a simple question, to first order we interpret the PSD breaks at each electromagnetic frequency, $\nu$, as tracing $z_\nu$. To transform between the two quantities, we employ the prescription,  $f_{\rm \,break}=\frac{\beta c \delta}{z_\nu}$ (see Table~\ref{table:psdfits})\footnote{Under the assumption of a conical jet, the distance downstream and the jet cross-section will be linked. Therefore, we also ran an alternate version of our MCMC modelling, where we assume the PSD breaks are explicitly linked to the jet cross-section instead; $f_{\rm \,break}=\beta_{\rm exp} c \delta/2 z_\nu\tan\phi$. This alternate prescription produced best-fit parameters that were very similar to the original modelling runs, attesting to the robustness of our modelling results.}.

Lastly, an observed time lag between two electromagnetic frequency bands can be written as follows,

\begin{equation}
\tau_{\rm lag}=z_{\rm norm}\left(\frac{1}{\nu_{\rm low}}-\frac{1}{\nu_{\rm high}}\right)\frac{(1-\beta\cos{i})}{\beta c}\,\,\,\,{\rm s}
\label{eq:lag}
\end{equation}
where $\nu_{\rm low}$ and $\nu_{\rm high}$ represent the lower and higher electromagnetic frequency bands being compared, the $(1-\beta\cos{i})$ term represents a correction due to the transverse Doppler effect\footnote{The transverse Doppler effect describes the situation where the observed interval between the reception of two photons is smaller than the emission interval.}, and $z_{\rm norm}=z_\nu\nu$ is the constant displayed on the right side of Equation~\ref{eq:z}.

 \renewcommand\tabcolsep{6pt}
 \begin{table}
\caption{Jet modelling results}\quad
\centering
\begin{tabular}{lcc}
 \hline\hline
 {\bf Parameter}&{\bf Best-fit Value }&{\bf Model Status}$^\ddagger$\\[0.15cm]
  \hline
 Jet power ($\log W$; ${\rm erg\,s}^{-1}$)$^\dagger$&$36.98^{+0.27}_{-0.35}$ &Free\\[0.15cm]
  Jet speed ($\Gamma$)$^\star$ & $6.81^{+1.06}_{-1.15}$& Free\\[0.15cm]
  Jet opening angle ($\phi$; deg)&$0.45^{+0.13}_{-0.11}$ & Free\\[0.15cm]
  Spectral index ($\alpha$)& $0.25^{+0.02}_{-0.02}$& Free\\[0.15cm]
  Distance ($D$; kpc)& $2.96\pm 0.33$ &Known\\[0.15cm]
  Inclination angle ($i$; deg)&$63\pm3$  & Known \\[0.15cm]
  Filling factor ($f$)& 1& Fixed\\[0.15cm]
  Equipartition fraction ($\xi_B$)&1 &Fixed \\[0.15cm]
  Proton contribution ($\xi_p$)&0 &Fixed \\[0.15cm]
 \hline
\end{tabular}
\begin{flushleft}
$^\dagger$ Given the posterior distribution of $W$, we estimate the distribution of the total power ($P=W\Gamma$), by performing Monte Carlo simulations sampling from the $W$ and $\Gamma$ distributions 10000 times to yield, $\log P=37.79^{+0.31}_{-0.38}$.\\

$^\star$ Given the posterior distribution of $\Gamma$, we estimate the distribution of the corresponding bulk jet speeds ($\beta$), by performing Monte Carlo simulations sampling from the $\Gamma$ distribution 10000 times to yield, $\beta=0.98^{+0.01}_{-0.01}$.\\
$^\ddagger$ This column indicates whether a parameter was left free, fixed, or known from an independent study (in this case we sampled from the known distribution in the fitting process).
\end{flushleft}
\label{table:jetmodel}
\end{table}
\renewcommand\tabcolsep{6pt}

Overall, Equations~\ref{eq:Fnu} through \ref{eq:lag} can allow us to predict average flux densities, PSD breaks, and lags. However, if we fit these data dimensions separately, the jet parameters in the model ($W$, $\beta$, $\phi$, $i$, $D$) will be highly degenerate. Alternatively, if we simultaneously fit all the data, tying the parameters between the data dimensions, we can help to break this degeneracy. Further, in the case of MAXI J1820+070, there exist independent constraints on the the distance ($D$) and inclination angle ($i$) from radio parallax measurements \citep{atri20,bright20}, which also help to reduce degeneracy in the model.

To solve for the jet power, opening angle, and speed, we use a MCMC algorithm to simultaneously fit the average jet spectrum (Table~\ref{table:avgfluxes}), PSD breaks (Table~\ref{table:psdfits}), radio-radio lags (Table~\ref{table:ccfval}), and sub-mm-radio lags (Table~\ref{table:alma_lag}). In this fitting process, we independently sample from the known distance ($D=2.96\pm0.33$ kpc) and inclination angle ($i=63\pm3^{\circ}$) distributions, fix the filling factor\footnote{Note that setting a lower filling factor value mainly affects the jet power parameter (where lower $f$ values lead to higher power estimates), while the other parameters do not change as significantly.} ($f=1$) and particle properties ($\xi_B=1$, $\xi_p=0$), and leave the jet power ($W$), speed ($\Gamma$), opening angle ($\phi$), and spectral index ($\alpha$) as free parameters. We use wide uniform priors for all of our free parameters. Additionally, we take steps to make the model more computationally efficient, by choosing to fit for the bulk Lorentz factor $\Gamma$ rather than $\beta$ (thereby avoiding hard boundaries for the speed parameter), and choosing to fit for $\log W$ rather than $W$ (thereby avoiding very large numbers).
The best fit result is taken as the median of the resulting posterior distributions, and the uncertainties are reported as the range between the median and the 15th percentile (-), and the 85th percentile and the median (+), corresponding approximately to $1 \sigma$ errors.
Table~\ref{table:jetmodel} shows the best-fit parameters, Figure~\ref{fig:jetmod} displays the best-fit model overlaid on the data, and corner plots displaying the posterior distributions and two-parameter correlations can be found in Figure~\ref{fig:corner}.
The \citet{blandford79} jet model can reproduce the different dimensions of our data quite well. 
Modelling our data with more complex jet models (e.g., \citealt{mal18}) will be considered in future work.

\begin{figure}
\begin{center}
  \includegraphics[width=0.45\textwidth]{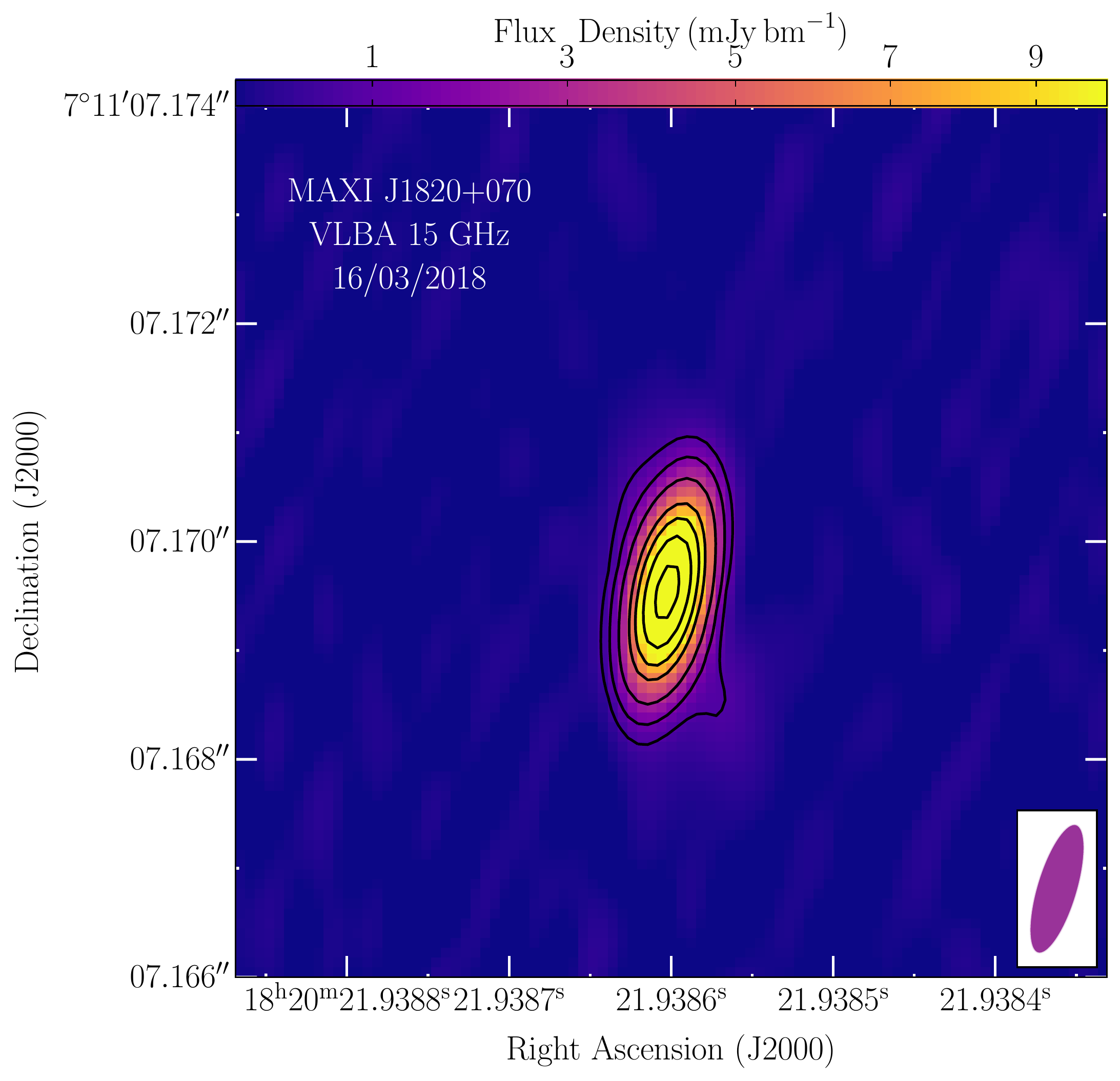}\\
 \caption{\label{fig:jetvlba} VLBA image of the marginally resolved MAXI J1820+070 jet at 15 GHz, created with observations taken on 2018 March 16 (Project Code: BM467). Contour levels are $2^{n/2}\,\times$ the rms noise level of $0.1\,{\rm mJy\, bm}^{-1}$ ($n=6,8,10,12,13,14,15$). The VLBA beam is shown in the {\it bottom right} corner. From this image, we measure a jet elongation of $l=0.52\pm0.02$ mas (${\rm PA}=25.9\pm2.8$ deg).
 }
\end{center}
 \end{figure}

As a further test of the accuracy of the model, we can compare the predictions of the best-fit model to a jet elongation measurement made independently with Very Long Baseline Array (VLBA) imaging of the MAXI J1820+070 jet. We imaged MAXI J1820+070 with the VLBA\footnote{For details on these VLBA observations (Project Code: BM467) and the data reduction process please see \citealt{atri20}.} on 2018 March 16 (about a month prior to the fast timing observations; see Figure~\ref{fig:jetvlba}). Through fitting the source with a Gaussian in the image plane, we measure the jet size scale in the plane of the sky to be $l=0.52\pm0.02$ mas at 15 GHz. To transform this measurement into a physical distance, we can use the following relation,
\begin{equation}
z_{\rm vlba}=(1.49\times10^{13})\frac{l_{\rm mas} D_{\rm kpc}}{\sin i}\,\,\,\, {\rm cm}
\end{equation}
Substituting in the known values of $D=2.96$ kpc and $i=63$ deg, results in $z_{\rm vlba}=25.7\times10^{12}$ cm, which is remarkably close to our best-fit model prediction (cyan X in Figure~\ref{fig:jetmod} \textit{middle}).


\section{Discussion}
\label{sec:discuss}
In this work, we have discovered highly variable, correlated multi-band emission from the BHXB MAXI J1820+070. Using Fourier and cross-correlation analyses, we measured the variability characteristics of the emission, and modelled these variability characteristics to directly estimate jet properties (e.g., power, speed, geometry, size scale). In the following sections, we discuss these jet properties, putting them into context with previous studies of MAXI J1820+070, as well as other BHXB systems. Additionally, we highlight the technical capabilities and instrumental advancements needed to push these types of BHXB spectral timing studies forward.

\subsection{Jet properties}
\label{sec:jprop}

\subsubsection{Jet size-scales}
In our Fourier analysis of the emission from MAXI J1820+070 (presented in Figures~\ref{fig:psd} and \ref{fig:psdX}), we discovered a clear evolution in the shape of the PSDs with electromagnetic frequency band. In particular, the PSD break frequency appears to scale inversely with electromagnetic frequency band through the jet-emitting bands (radio--optical), before leveling off into a plateau as we reach the X-ray band (see Figure~\ref{fig:psd_metrics}). This trend matches the relationship we expect to see between the downstream distance of the emitting region from the black hole and electromagnetic frequency band.
Thus measuring the PSD break frequency at several bands, has allowed us to, for the first time, map out the jet size scale with electromagnetic frequency (see Table~\ref{table:psdfits}; note that \citealt{vinc19} have also previously suggested a tentative connection between high Fourier frequency IR PSD features and the jet size scale for GX 339--4).

Our jet size scale predictions show remarkable consistency with previous work on MAXI J1820+070. For example, \citet{pai19} presented optical ($g_s$ band; equivalent to $6.4\times10^5$ GHz) observations of MAXI J1820+070, taken 5 days after our observations. The $g_s$ band PSD displays a break $\sim5-50$ Hz, whereas our model predicts the PSD break at 12.4 Hz (although we note that there may be other sources of optical emission from the system; \citealt{vel19,kos20}).
Additionally, \citet{marks20} present an upper limit of $<0.1$ mas for the size scale of the infrared emitting region (in the plane of the sky) in MAXI J1820+070, from direct imaging with the GRAVITY instrument on the VLT Interferometer (observations taken between 2018 May 31 - June 1, while the system was still in the hard state). This measurement corresponds to physical scales of $\lesssim10^{12}$ cm (assuming a distance of 2.96 kpc), in agreement with the infrared emission region measurements reported here. 
However, we do notice some deviations between our best-fit model and the data (see Figure~\ref{fig:jetmod}), where the model can over/under-predict the $z_\nu$ and lags. These deviations may be indicative of acceleration occurring in the jet flow, or alternatively a breakdown of the expected (linear) size scale to electromagnetic frequency relation farther out in the jet flow. 
Such a breakdown could possibly suggest a non-conical jet geometry. Interestingly, in a previous radio timing study of Cygnus X--1 \citep{teta19}, a similar effect was observed, where a shallower\footnote{We note that the Cygnus X--1 system contains a high-mass donor star with a strong stellar wind, which is known to at least partially absorb radio signals \citep{poo99cyg,bro02}. Therefore, it is a possibility that this shallower relation may be related to the wind absorption effects.} $z_\nu\propto\nu^{-0.4}$ relation was needed to match the time lag measurements at lower (S Band; 2--4 GHz) electromagnetic frequencies.
Further, possible evidence for a non-conical jet geometry has been reported recently for neutron star XB 4U 0614+091 \citep{marino20}.

\subsubsection{Jet speed}
\label{sec:speed_discuss}
Through modelling the jet timing characteristics, we have found that MAXI J1820+070 houses a highly relativistic compact jet. In particular, we estimate a bulk Lorentz factor of $6.81^{+1.06}_{-1.15}$. This far exceeds what was estimated\footnote{Note that the transient and compact jet speeds are measured with different methods here. Further, \cite{fend03}  have shown that for any significantly relativistic source (which must be located close to $d_{\rm max}$), the Lorentz factor varies rapidly with distance, and thus could be a lot higher.} for the transient jet ejections that occurred later on in the outburst ($\Gamma\sim2.2$; \citealt{atri20,bright20}), {and is also higher than compact jet speeds suggested by the two other BHXBs with direct speed measurements}; Cygnus X--1 ($\Gamma=2.6$; \citealt{teta19}) and GX 339--4 ($\Gamma>2$; \citealt{cas10}). It is thought that compact jet speed increases as the outburst progresses through the rising hard state \citep{vadawale2003,fbg04,fenhombel09}. MAXI J1820+070 rose quickly ($\sim10$ days) through this rising hard state \citep{meg18,kaj19}, in turn reaching a high luminosity/fast-jet state very early on in the outburst, where it remained for several months before transitioning to the soft state (when the compact jet is quenched). As our observations were taken when the source was in this high luminosity/fast-jet state, this could explain the higher jet speed measurement here, and suggests we have sampled the jet speed near the high end of its distribution in this outburst.

\begin{figure}
\begin{center}
  \includegraphics[width=0.45\textwidth]{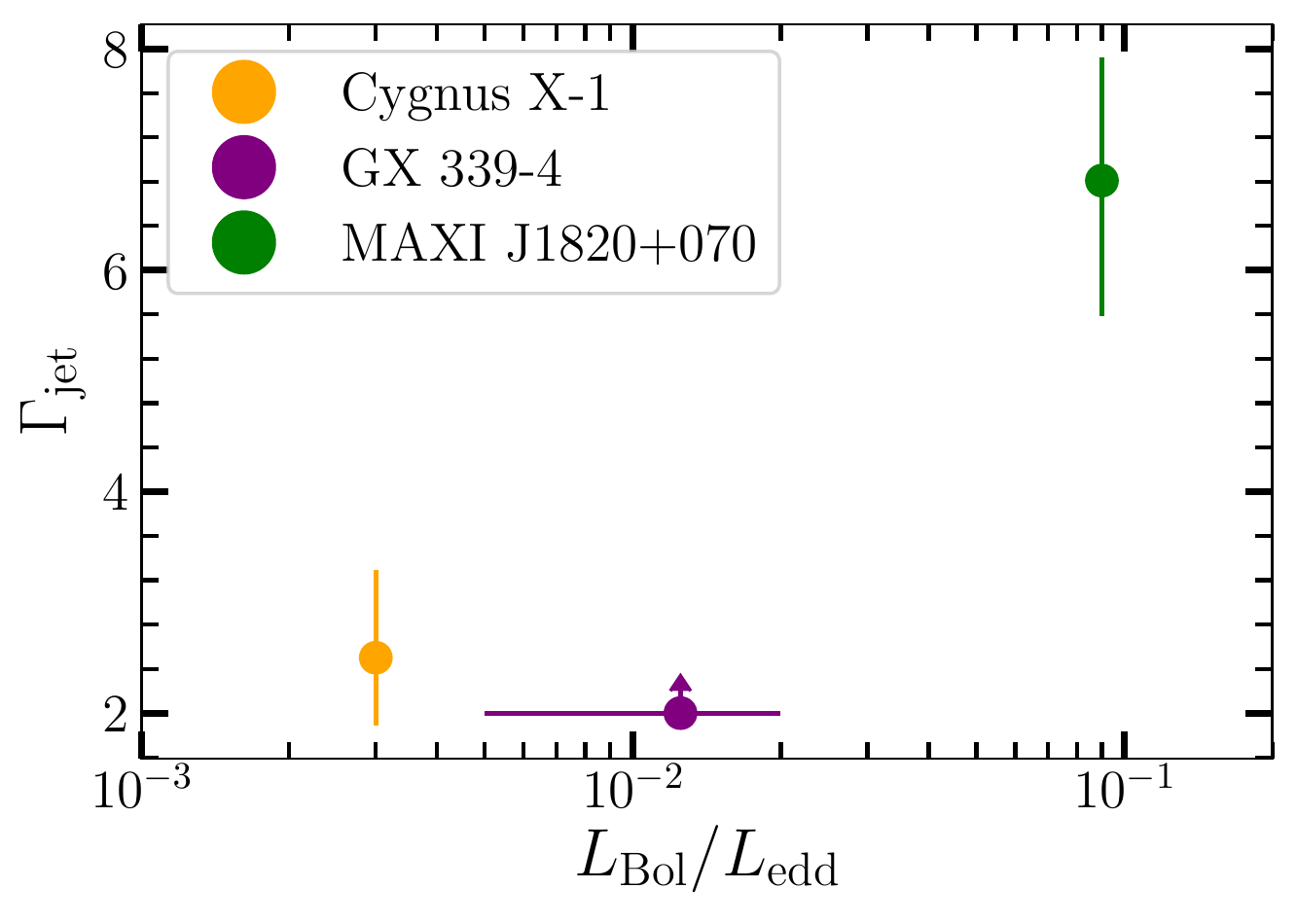}\\
 \caption{\label{fig:gamma_ledd} Comparison between jet bulk Lorentz factors and Eddington fractions for the three BHXBs with compact jet speed constraints; Cygnus X--1 \citep{teta19}, GX 339--4 \citep{cas10}, MAXI J1820+070 (this work; \citealt{meg18,kaj19}). The horizontal error bars for the Eddington fraction of GX 339--4 represent the range of black hole masses estimated for this source (see \S\ref{sec:speed_discuss} for details). Given the small sample of sources with jet speed constraints so far, it is difficult to definitively determine if we observe direct evidence for the jet speed being correlated with the Eddington fraction.}
\end{center}
 \end{figure}

Considering the three BHXBs with compact jet speed constraints (GX 339--4, Cygnus X--1, and MAXI J1820+070; \citealt{cas10,teta19}), we can compare the jet bulk Lorentz factors to the Eddington fraction when the jet speed measurements were made (see Figure~\ref{fig:gamma_ledd}). To estimate the Eddington luminosity for each source, we use $M_{\rm BH}=21.2 M_\odot$ for Cygnus X-1 \citep{mj21}, $M_{\rm BH}=2.3-9.5 M_\odot$ for GX 339--4 \citep{cs16}, and $M_{\rm BH}=8.48 M_\odot$ for MAXI J1820+070 \citep{torr19,torr20}. To estimate bolometric X-ray luminosity, we use $D=2.2$ kpc for Cygnus X-1 \citep{mj21}, $D=8$ kpc for GX 339--4 \citep{cs16}, and $D=2.96$ kpc for MAXI J1820+070 \citep{atri20}, as well as
the conversion $F_{\rm Bol}\sim 5\times F_{2-10{\rm keV}}$ for BHXBs in the hard accretion state \citep{migfen06}.
As the source sample is small, it is difficult to definitively determine whether jet speed increases with Eddington fraction, in line with past predictions \citep{vadawale2003,fbg04,fenhombel09}, measurements of two tracks in the radio/X-ray correlation for BHXBs \citep{rus15}, and more recent modelling \citep{pea19}. {However, Figure~\ref{fig:gamma_ledd} does hint at the presence of a positive correlation between jet speed and Eddington fraction in BHXBs.}

\subsubsection{Jet power}

Our modelling has also allowed us to estimate the jet kinetic power, where we find $P=(6.2^{+6.4}_{-3.6})\times10^{37}\,{\rm erg\,s}^{-1}$. Upon comparing this jet power estimate to X-ray studies closest in time to our observations in the hard state of this outburst, we find that the power output of the MAXI J1820+070 jet is a significant fraction of the bolometric X-ray luminosity ($f_{\rm x}={P_{\rm jet}}/{L_{1-100{\rm keV}}}\sim0.6$, where $L_{1-100{\rm keV}}\sim1.0\times10^{38}\,{\rm erg\,s}^{-1}$; \citealt{meg18,kaj19}), and similar to the power estimated to be carried in the transient jet ejecta ($\sim4\times10^{37} {\rm erg\,s}^{-1}$; \citealt{bright20})\footnote{Note that this power estimate assumes that the jet ejecta are associated with an optically thin radio flare, and were launched over a time period $\lesssim 6.7$ hours (equivalent to the rise-time of this flare).}. The discovery of such a high-power compact jet (especially when compared to the X-ray output) in MAXI J1820+070 is reminiscent of the jet launched from the BHXB Cygnus X--1, which is known to carry energies on par ($f_{\rm x}\sim0.06-1$) with the bolometric X-ray luminosity in the system, and has carved out a large pc-scale cavity in the surrounding ISM as a result \citep{gallo05}. In the several-month period that MAXI J1820+070 spent in the hard state, with its compact jet turned on, it would have deposited a large amount of energy into the local ISM (over 4 months, $\sim6\times10^{44}\,{\rm erg}$), and thus we might expect to observe a similar feedback effect here, where the jet may have carved out an ISM cavity (if MAXI J1820+070 is located in a dense enough environment and not moving supersonically relative to the ISM; e.g., the Cygnus X--1 jet is propagating through the tail of an HII region; \citealt{gallo05}). This theory is consistent with the discovery of X-ray hot-spots later on in the MAXI J1820+070 outburst, which could be produced by shocks between the edge of this ISM cavity and the jet ejecta \citep{esp20}.

\subsubsection{Jet geometry}
We were also able to constrain the jet opening angle in MAXI J1820+070, finding $\phi=0.45$ deg. While the majority of opening angle constraints in the BHXB population are upper limits \citep{millerj06}, the MAXI J1820+070 opening angle is one of the narrowest opening angles measured to date (although see \citealt{mal18}, which predicts opening angles as low as 0.05 degrees in GX 339--4). This small opening angle could suggest a highly confined jet. While it is difficult to pinpoint the confinement mechanism, given that a strong accretion disc wind was detected during the hard state of this outburst \citep{md19}, it is possible that this wind could have played a significant role in inhibiting the transverse expansion of the jet.

\subsubsection{Jet composition}
The high jet power and bulk jet speed that we have found for MAXI J1820+070 can allow us to place constraints on the composition of the jet (i.e., proton content). In particular, if there are too many protons (positron to proton ratio is too low), then the bulk kinetic power in the jet will become physically unreasonable.

We know that the jet kinetic power can not be dramatically higher than the accretion power near the hard to soft accretion state transition, as this would suggest a discontinuity in the mass accretion rate across the transition, which is ruled out observationally by the smoothness of X-ray light curves across the transition for many BHXB sources \citep{macca05}.
In \S\ref{sec:model}, we have shown that for a purely leptonic (pair-dominated) jet, the kinetic power is already a significant fraction of the accretion power ($\sim0.6 L_{{\rm acc}}$). In fact, if the jet were proton-dominated (no pairs), then the kinetic power would be $\frac{m_p}{\gamma_{\rm min}m_e}\sim 30$ times the purely leptonic jet power (where the low energy cutoff of the electron energy distribution $\gamma_{\rm min}\sim 70 (\nu/8.4\,{\rm GHz})^{0.5}(p/1\,{\rm erg\,cm}^{-3})$, and we set $\nu=5.25$ GHz, corresponding to our lowest sampled electromagnetic frequency; \citealt{heinz06}). Therefore, a proton-dominated jet would have a kinetic power that exceeds the accretion power. If we use the accretion power as an upper limit, we can place quantitative constraints on the proton to electron ratio ($N_p/N_e$) in the jet. Including the proton content, the total kinetic power can be written as \citep{heinz06},
\begin{equation}
    W_{\rm total}= W + 2\rho c^3 \Gamma\beta (\Gamma-1)\pi \phi^2z^2
    \label{eq:Wtot}
\end{equation}
where $W$ is the power in the purely leptonic jet estimated from our modelling (Equation~\ref{eq:W}, Table~\ref{table:jetmodel}). Through rearranging Equation~\ref{eq:Wtot}, and setting $W_{\rm total}=L_{\rm acc}$, we find a density $\rho=6.5\times10^{-20}\,{\rm g\, cm}^{-3}$, which corresponds to $N_p/N_e\sim0.6$. Therefore, this approach suggests that the jet in MAXI J1820+070 can not be proton dominated.

\subsubsection{Strength of the counter-jet signal}
Lastly, given our measurements of the jet speed, inclination angle, and spectral index ($\beta=0.98$, $i=63^\circ$, and $\alpha=0.25$; see Table~\ref{table:jetmodel}), we can constrain the strength of the signal from the counter-jet (the portion of the bi-polar jet travelling away from us). We estimate a ratio of the flux densities between the approaching and receding jets for MAXI J1820+070 of $\frac{F_{\rm app}}{F_{\rm rec}}=\left(\frac{1+\beta\cos i}{1-\beta\cos i}\right)^{2-\alpha}=5.3$. This suggests that the counter-jet signal is not negligible. Analyzing the effect of the counter-jet signal on our timing analysis will be explored in future work.

\subsection{Designing future spectral timing experiments}
\label{sec:future}

While spectral timing studies of BHXBs continue to provide new insights into accretion driven jets, these studies, especially in the lower electromagnetic frequency bands, are very much in their infancy. Therefore, it is important that we continue to evaluate how these experiments can be improved, and how we can design future observing campaigns to overcome any challenges we currently face. 

In this study, we discovered that a significant smoothing effect between the sub-mm and radio signals can make measuring a sub-mm-radio lag infeasible with the CCF method. This issue highlights the need to simultaneously sample more closely spaced frequency bands in between the two regimes (30--100 GHz). At present, this would entail adding more instruments to a campaign (and the added difficulty that comes with synchronizing more telescopes). But with planned next generation instruments, like the next generation VLA \citep{ngvla}, we could observe these intermediate bands with one instrument, and in turn make these types of observations much more feasible. Additionally, if we could have the ability to implement a VLA-type sub-array technique with ALMA, this could also help sample a wider range of bands simultaneously.

Further, the majority of spectral-timing studies (including this one) have been one-shot observations, and thus unable to probe how jet variability evolves during outburst. The formalism we have developed here, to measure jet properties using timing characteristics alone, could allow us to map out for the first time how quantities such as jet power and speed change throughout an outburst. While repeating this spectral timing experiment several times throughout an outburst would require significant time commitment from several observatories, implementing dedicated large multi-semester observing programs presents a viable option in this respect (e.g., JCMT large program PITCH-BLACK \footnote{\url{https://www.eaobservatory.org/jcmt/science/large-programs/pitch-black/}}; Tetarenko et al. in prep.).

Lastly, this work has shown the importance of combining timing studies at lower electromagnetic frequency bands (radio, sub-mm) with those at higher electromagnetic frequency bands (OIR), to connect variability properties across different scales in the jet/accretion flow. Therefore, the inauguration of highly sensitive next generation instruments, like the Square Kilometre Array (SKA) and James Webb Space Telescope (JWST), present exciting prospects for continuing to develop multi-wavelength spectral timing experiments.

\section{Summary}
\label{sec:sum}

In this paper, we present high time resolution multi-wavelength measurements of the BHXB MAXI J1820+070 during the hard state of its 2018--2019 outburst. These observations were taken with the VLA, ALMA, VLT HAWK-I, NTT ULTRACAM, NICER, and XMM-Newton, sampling a total of ten different electromagnetic frequency bands simultaneously. We find that the emission from MAXI J1820+070 is highly variable, showing multiple structured flaring events over a 7-hour observation period.

To characterize the variability we observe, we use a combination of cross-correlation and Fourier analyses. Through these analyses, we discovered that the emission is highly correlated between the different electromagnetic frequency bands, showing clear time-lags ranging from minutes between the radio/sub-mm bands to hundreds of ms between the X-ray/optical bands.
A Fourier analysis of the emission revealed a clear trend in the PSD break frequency with electromagnetic frequency band, which has allowed us to map out the jet size scale for the first time in a BHXB.
Additionally, through modelling the multi-band variability properties in MAXI J1820+070 with a Bayesian formalism, we directly measured jet speed, geometry, and energetics, finding a highly relativistic ($\Gamma=6.81^{+1.06}_{-1.15}$), confined ($\phi=0.45^{+0.13}_{-0.11}$ deg) jet, which carries a significant amount of energy away from the black hole ($\log P=37.79^{+0.31}_{-0.38}\,{\rm erg\,s}^{-1}$, equivalent to $\sim0.6 \, L_{1-100{\rm keV}}$). We use this high jet power and bulk jet speed to place constraints on the jet composition, finding that the jet in MAXI J1820+070 can not be proton dominated. Lastly, we put constraints on the magnetic field strength in the jet base region of $B>6\times10^3$ G.

Overall, this work demonstrates that it is possible to accurately measure key jet properties using only time-domain measurements. To take full advantage of these time-domain tools, it is essential that we continue to develop these spectral timing techniques, and repeat these experiments throughout different outburst states and across different BHXB systems.

\section*{Acknowledgements}
The authors thank the anonymous referee for the time and effort put into reviewing this manuscript and the helpful feedback they provided, especially in these complicated times during the COVID-19 pandemic.
We offer a special thanks to the NRAO for granting our DDT request for the VLA observations presented in this paper.
The authors also wish to thank Anna Pala and Kaustubh Rajwade for their efforts obtaining and processing the ULTRACAM data used in this work.
AJT additionally thanks Bailey Tetarenko for discussions on bolometric X-ray luminosity calculations.
JCAMJ is the recipient of an Australian Research Council Future Fellowship (FT140101082), funded by the Australian government. GRS is supported by an Natural Sciences and Engineering Research Council of Canada Discovery Grant (NSERC; RGPIN-06569-2016). JAP is supported in part by a University of Southampton Central VC Scholarship. JAP and PG acknowledge STFC and a UGC-UKIERI Thematic partnership for support. VSD, TRM, and ULTRACAM are
supported by the STFC. This paper makes use of the following ALMA data: ADS/JAO.ALMA\#2017.1.01103.T. ALMA is a partnership of ESO (representing its member states), NSF (USA) and NINS (Japan), together with NRC (Canada), MOST and ASIAA (Taiwan), and KASI (Republic of Korea), in cooperation with the Republic of Chile. The Joint ALMA Observatory is operated by ESO, AUI/NRAO and NAOJ. The National Radio Astronomy Observatory is a facility of the National Science Foundation operated under cooperative agreement by Associated Universities, Inc. We acknowledge use of the \textsc{stingray} and \textsc{corner} \textsc{python} packages, as well as the ZDCF codes of \citet{alex97}, for this work.

\section*{Data Availability}
The radio/sub-mm data presented in this work are available in the ALMA (\url{http://almascience.nrao.edu/aq/}) and NRAO (\url{https://archive.nrao.edu/archive/}) online data archives. Raw HAWK-I cubes can be found in the ESO online archive (\url{http://archive.eso.org/eso/eso_archive_main.html}). The ULTRACAM data can be made available upon reasonable request to the authors. XMM-Newton and NICER event files can be downloaded from their respective online archive repositories (\url{http://nxsa.esac.esa.int/nxsa-web/#search} and \url{https://heasarc.gsfc.nasa.gov/cgi-bin/W3Browse/w3browse.pl}).




\bibliography{ABrefList}



\appendix

\section{Radio and sub-mm calibrator light curves}
 \label{sec:ap_cal}
\begin{figure*}
\begin{center}
  \includegraphics[width=0.9\textwidth]{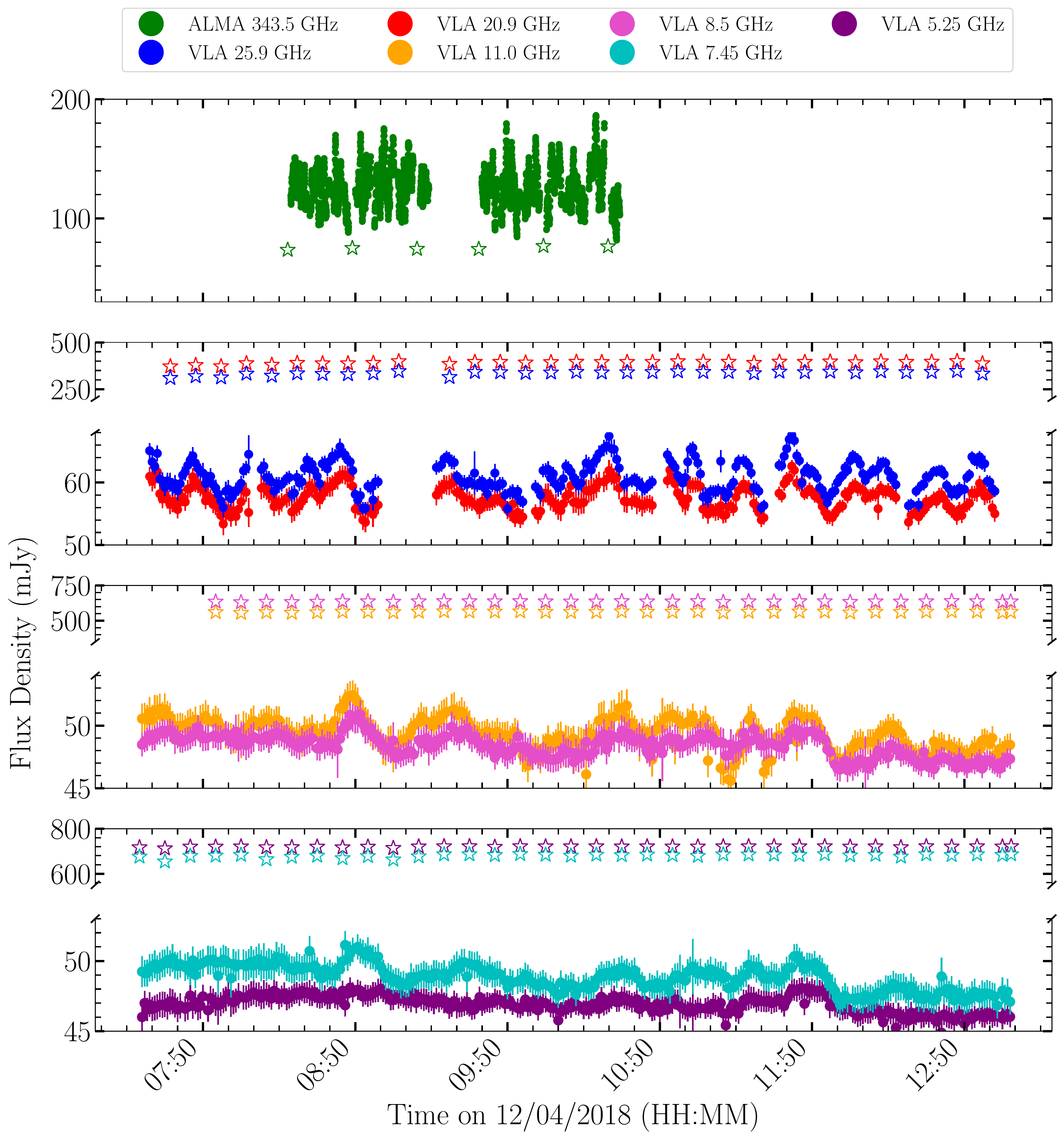}\\
 \caption{\label{fig:lc_cal}Multi-band radio and sub-mm light curves of MAXI J1820+070 and our calibrator check sources (J1832+0731 for 343.5 GHz and J1824+1044 for 5.25--25.9 GHz). The panels from \textit{top} to \textit{bottom} show light curves for progressively decreasing electromagnetic frequency bands (as indicated by the legend). All of the check sources are plotted as star symbols. As all of the check source observations show relatively constant flux densities over the course of the observations, the variations we observe in MAXI J1820+070 are most likely intrinsic to the source, and not the result of atmospheric or instrumental effects.}
\end{center}
 \end{figure*}

Given the flux variability that we detected in our radio and sub-mm light curves of MAXI J1820+070, we wanted to ensure that the variations observed represent intrinsic source variations, rather than atmospheric or instrumental effects. Therefore, we created time resolved light curves of check sources, which are bright calibrators that are treated as science targets in the data reduction (see Figure~\ref{fig:lc_cal}). While the ALMA observations were set up specifically with a check source that differed from the phase calibrator (J1832+0731), the VLA observations did not have a specific check source in the observational setup. Therefore, to mimic the check source in our VLA data, we re-ran the reduction, treating every other phase calibrator (J1824+1044) scan as a science target.
We find that all of the check sources display a relatively constant flux density throughout our observations, with any variations (<1\% of the average flux density) being a very small fraction of the variations we see in MAXI J1820+070. Based on these results, we are confident that our light curves of MAXI J1820+070 are an accurate representation of the rapidly changing intrinsic flux density of the source.


 \section{PSD white noise levels}
 \label{sec:ap_wn}
\begin{figure*}
\begin{center}
  \includegraphics[width=0.9\textwidth]{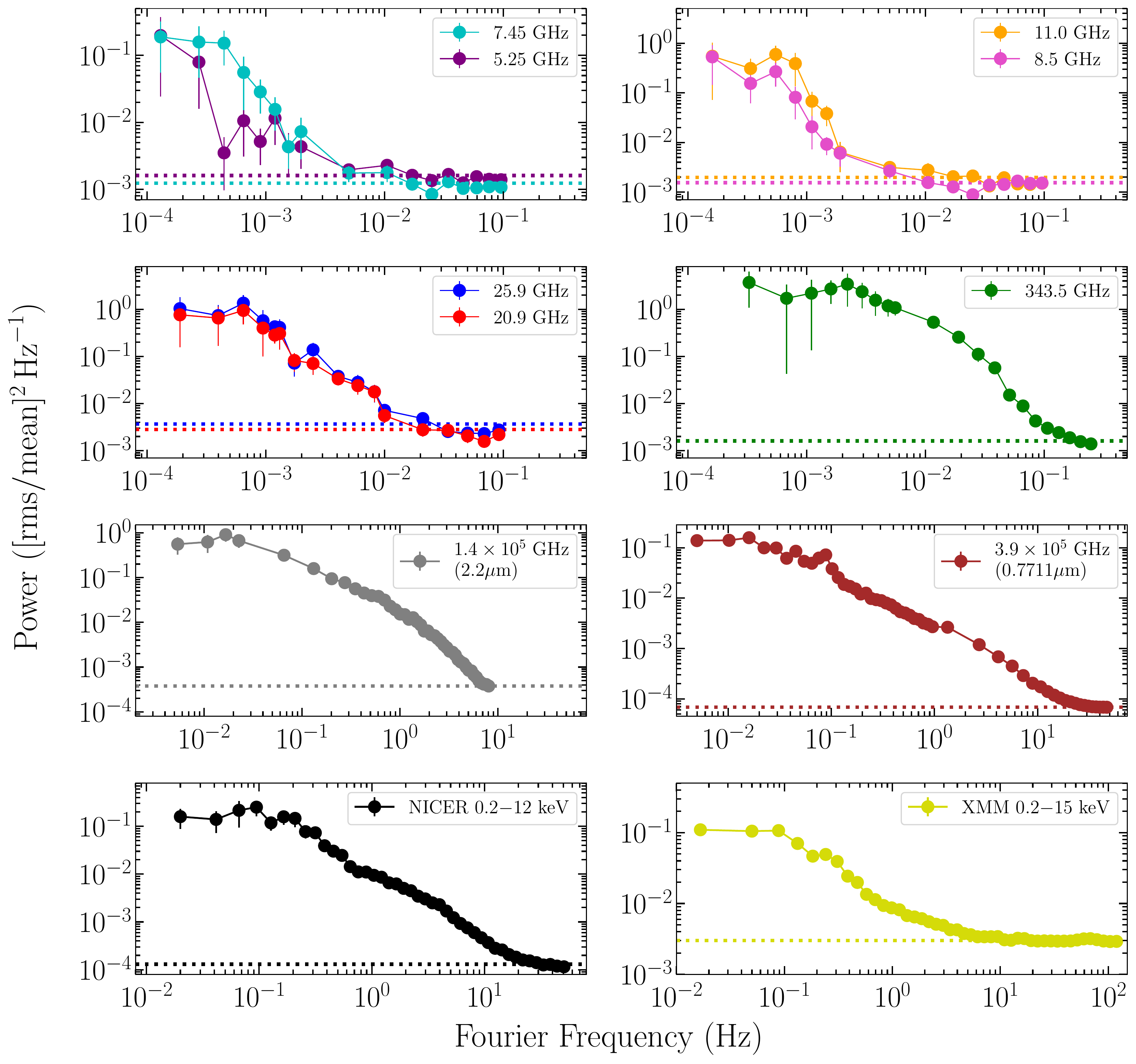}\\
 \caption{\label{fig:psd_wn} Fourier power spectra (PSDs) of emission from MAXI J1820+070, prior to white noise subtraction. The estimated white noise levels are indicated by the dotted lines in each panel. }
\end{center}
 \end{figure*}

In this section, we show the PSDs prior to white noise subtraction, and indicate the measured white noise levels (see Figure~\ref{fig:psd_wn}). 
We have estimated the white noise levels by fitting a constant to the highest Fourier frequencies.

Note that for the infrared/optical data ($2.2/0.7711\mu$m), we notice that this procedure tends to slightly overestimate the white noise levels. This is especially apparent in our PSD fits (see Figure~\ref{fig:psd_fits}), where we see a slight excess in the residuals at the highest Fourier frequencies, and also measure quite a steep PSD slope after the break. This effect is likely due to some non-flat component of the noise, potentially caused by the instrument readout. A precise estimate of this component goes beyond the scope of this work, but to ensure that this does not bias our key PSD measurement of the break frequency, we repeated the fit to the these PSDs prior to white noise subtraction (i.e., equivalent to adding a constant component into the model). We find that the best-fit PSD break in these secondary fits is consistent with the original fits within the $1\sigma$ uncertainties.

{Additionally, given the steep slopes after the PSD breaks that we measure, we opted to also investigate the effect of leakage caused by windowing on our PSD slopes. To test if this windowing effect is sufficiently strong to hide slopes steeper than 2 (in turn suggesting that incorrect white noise subtraction could be producing our steep slopes), we ran a set of simulations where we simulated light curves from a PDS with our measured breaks and slopes (and with the same windowing and sampling properties of our true light curves). We then calculated the PDS of these simulated light curves again, to check if the steep slope had disappeared. In our simulations, we find that we are still able to recover the steep slopes in the Fourier frequency range where we have significant power, before the white noise floor dominates, and thus this effect does not seem to be strong enough to fully hide the steeper slopes in our case.}

\section{PSD Fits}
 \label{sec:ap_psdfits}
\begin{figure*}
\begin{center}
  \includegraphics[width=0.8\textwidth]{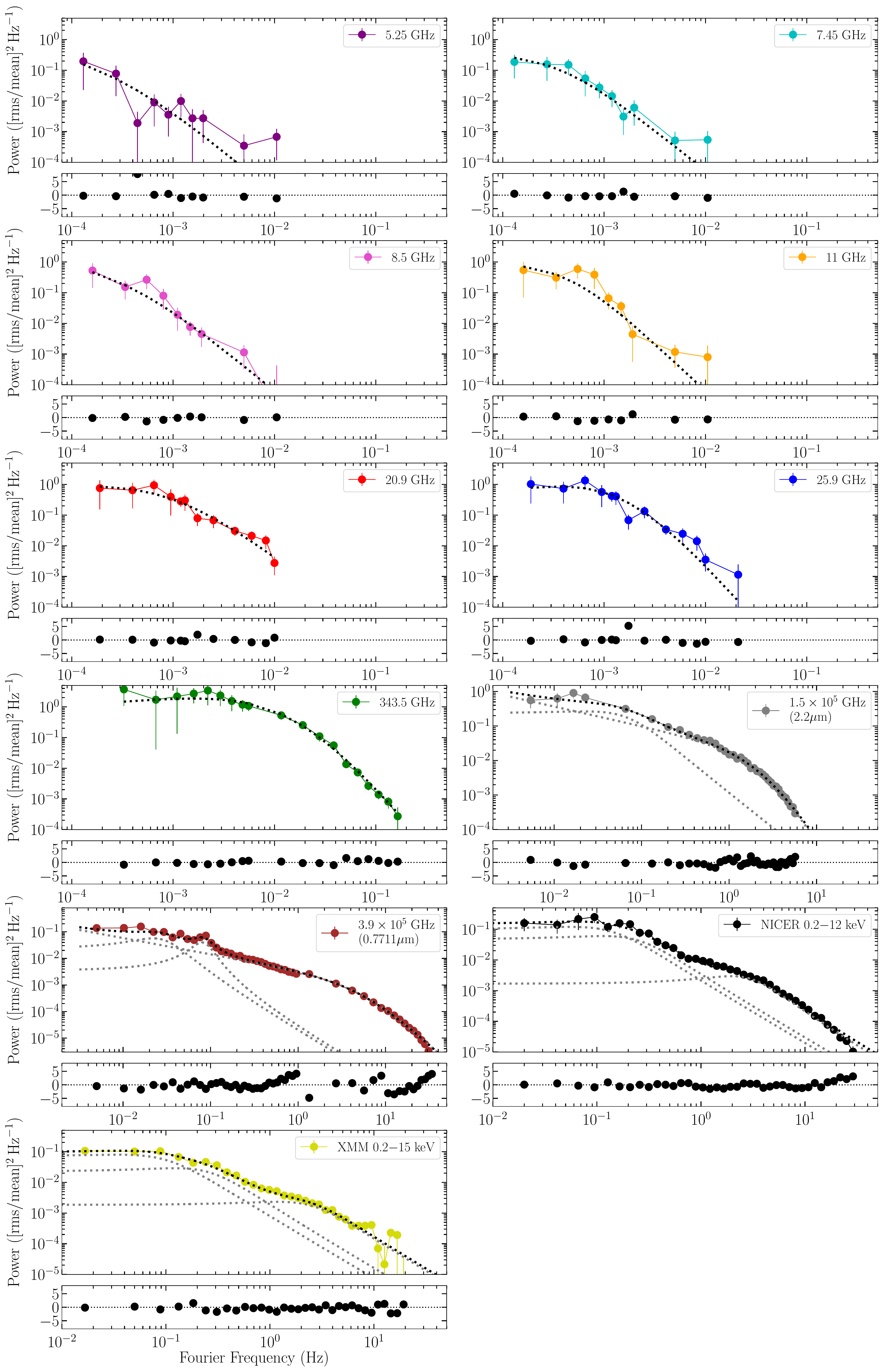}\\
 \caption{\label{fig:psd_fits}Fits to the MAXI J1820+070 Fourier power spectra (PSDs). In each panel, the PSDs and best-fit model (indicated by the dashed black line) are shown in the \textit{top} sub-panels, while the residuals for the fits (data-model/uncertainties) are shown in the \textit{bottom} sub-panels. The radio/sub-mm PSDs (5.25--343.5 GHz) are fit with a broken power-law, the infrared/optical ($2.2\mu$m/$0.7711\mu$m) PSDs are fit with a broken power-law for the higher Fourier frequencies + Lorentzian(s) for the lower Fourier frequencies, and the X-ray PSDs are fit with Lorentzians (see \S\ref{sec:psd} for details). In the cases where more than one component (e.g., broken power-law + Lorentzian) is needed to fit the total PSD, the individual components are displayed by gray dashed lines.}
\end{center}
 \end{figure*}

\begin{figure}
\begin{center}
  \includegraphics[width=0.48\textwidth]{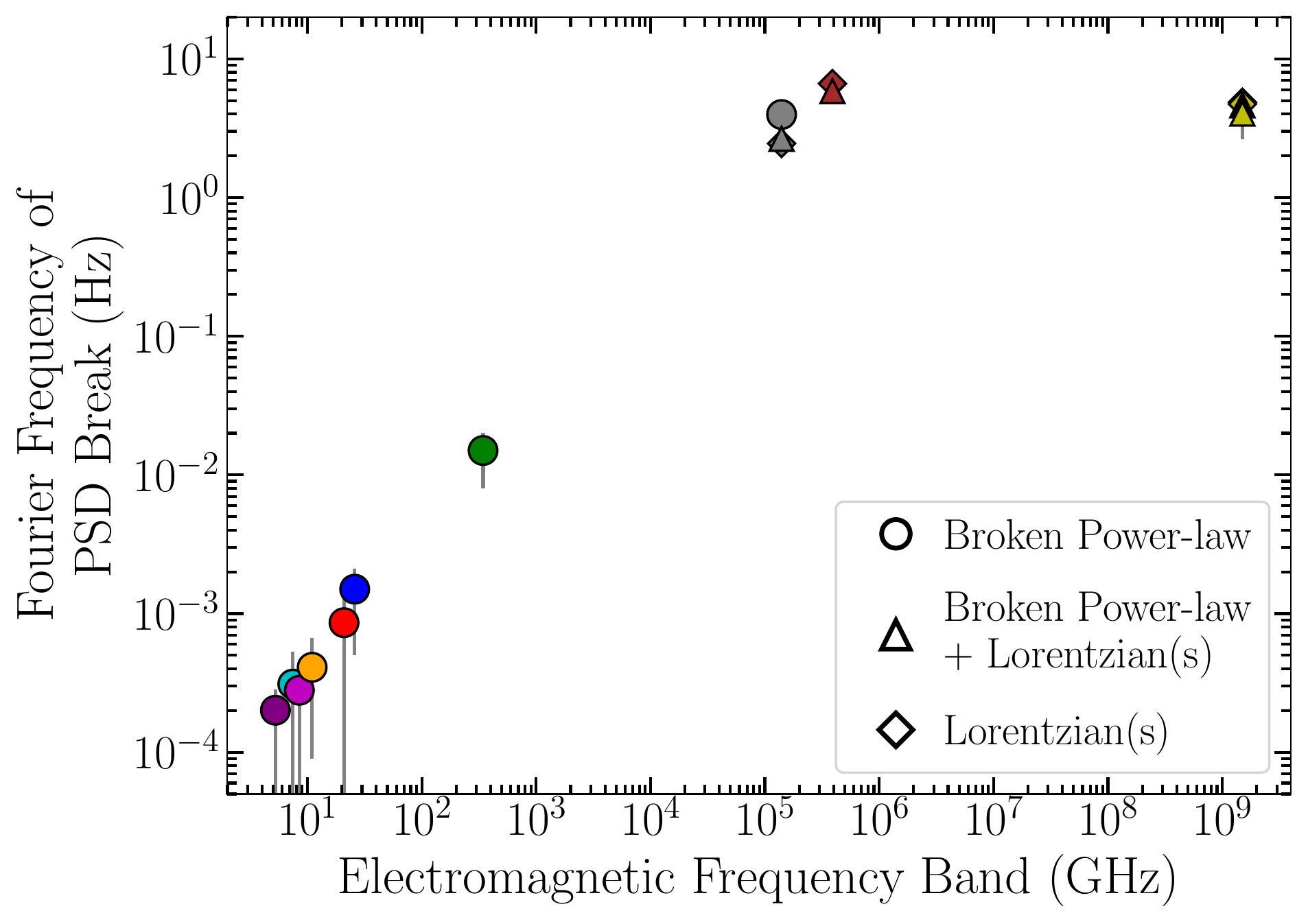}\\
 \caption{\label{fig:psd_metrics2} Fourier frequency of the PSD break inferred from using different models to fit the PSDs; broken power-law ({\it circles}), broken power-law for the higher Fourier frequencies + Lorentzian(s) for the lower Fourier frequencies ({\it triangles}), and strictly Lorentzian components ({\it diamonds}). {For the strictly Lorentzian component fits, we take the break frequency to be the ``characteristic frequency" defined in \citealt{belloni02} as $f_{\rm break}=\sqrt{\nu_0^2+\Delta^2}$, where $\nu_0$ is the central frequency and $\Delta$ is the FWHM of the highest Fourier frequency Lorentzian.}
 The colours of the data points correspond to the same colours of the electromagnetic frequency bands in Figures~\ref{fig:psd} \& \ref{fig:psdX}. Overall, we find that the PSD break measurements do not change drastically when different PSD models are used.}
\end{center}
 \end{figure}

In this section, we show the results of the final fits to our PSDs (Figure~\ref{fig:psd_fits}), as well as a comparison between the PSD break frequencies found using different PSD models (Figure~\ref{fig:psd_metrics2}).


\bsp	
\label{lastpage}
\end{document}